%% file: Main.tex
\documentclass[journal]{IEEEtran}

\usepackage{xcolor}
\usepackage{cite}
\usepackage{graphicx}
\usepackage[bb=dsserif]{mathalpha}
\usepackage{bm}
\usepackage{algorithm}
\usepackage{algorithmic}
\usepackage{array}
\usepackage{makecell}
\usepackage[short]{optidef}
\usepackage{amsthm}
\usepackage{amsmath}
\usepackage{amssymb}
\usepackage{mathrsfs}
\allowdisplaybreaks[4]

\usepackage[font=footnotesize]{subcaption}
\usepackage[font=footnotesize]{caption}
\usepackage{stfloats}
\newcommand\numeq[1]%
{\stackrel{\scriptscriptstyle(\mkern-1.5mu#1\mkern-1.5mu)}{=}}
\DeclareMathAlphabet\bfcal{OMS}{cmsy}{b}{n}
\include{notation_new}

\begin{document}
\title{A Novel Pilot Scheme for Uplink Channel Estimation for Sub-array Structured ELAA in XL-MIMO systems}

\author{Yumeng~Zhang~\IEEEmembership{Member,~IEEE}, Huayan~Guo,~\IEEEmembership{Member,~IEEE},  and Vincent K. N. Lau,~\IEEEmembership{Fellow,~IEEE}
\thanks{This work is supported by in part by the Research Grants Council of Hong
Kong under Project 16213119, in part by the Research Grants Council under
the Areas of Excellence scheme grant AoE/E-601/22-R and AoE/E-101/23-N,
in part by the National Natural Science Foundation of China under Grant
62101472 and  in part by the National Foreign Expert Project under Project H20240994. (Corresponding author: Huayan Guo; Vincent K. N. Lau.)}
\thanks{The authors are with the Department of Electronics and Computer Engineering, The Hong Kong University of Science and Technology, Hong Kong 999077, China (e-mail: eeyzhang@ust.hk,eeguohuayan@ust.hk, eeknlau@ust.hk).}
}

\maketitle
\begin{abstract}
This paper proposes a novel pilot scheme for multi-user uplink channel estimation in extra-large-scale massive MIMO (XL-MIMO) systems with extremely large aperture arrays (ELAA). The large aperture of ELAA introduces spatial non-stationarity, where far-apart users have significantly distinct visibility at the antennas, thereby reducing inter-user interference. This insight motivates our novel pilot scheme to group users with distinct visibility regions to share the same frequency subcarriers for channel estimation, so that more users can be served with reduced pilot overhead. Specifically, the proposed pilot scheme employs frequency-division multiplexing for inter-group channel estimation, while intra-group users -- benefiting from strong spatial orthogonality -- are distinguished by shifted cyclic codes, similar to code-division multiplexing.  Additionally, we introduce a sub-array structured ELAA, where each sub-array is a traditional MIMO array and treated as spatial stationary, while the distances between sub-arrays can be significantly larger to achieve an expanded aperture. The channel support for sub-arrays features clustered sparsity in the antenna-delay domain and is modeled by a 2-dimensional (2-D) Markov random field (MRF). Based on this, we propose a low-complexity channel estimation algorithm within a turbo Bayesian inference framework that incorporates the 2-D MRF prior model. Simulations show that the proposed scheme and algorithm allow the XL-MIMO system to support more users, and deliver superior channel estimation performance.

\end{abstract}


\IEEEpeerreviewmaketitle

\section{Introduction}
Extra-large-scale massive MIMO (XL-MIMO) has become a promising solution to support a large number of users while enhancing coverage in cellular systems by utilizing extremely large aperture arrays (ELAA) \cite{Martínez2014}.  Effective channel estimation is a prerequisite to unlock the full potential of XL-MIMO.  However, traditional channel estimation in massive MIMO suffers significant performance degradation due to the near-field effect, as users are generally located within the Rayleigh distance of the ELAA \cite{NonstationarityELAA,Flordelis,GaoX}. Additionally,  the pilot overhead increases 
with the number of users for uplink channel sounding, which considerably decreases the available payload within the limited coherent time. 

\begin{figure}[!t]
\centering
\includegraphics[width=.75\columnwidth]{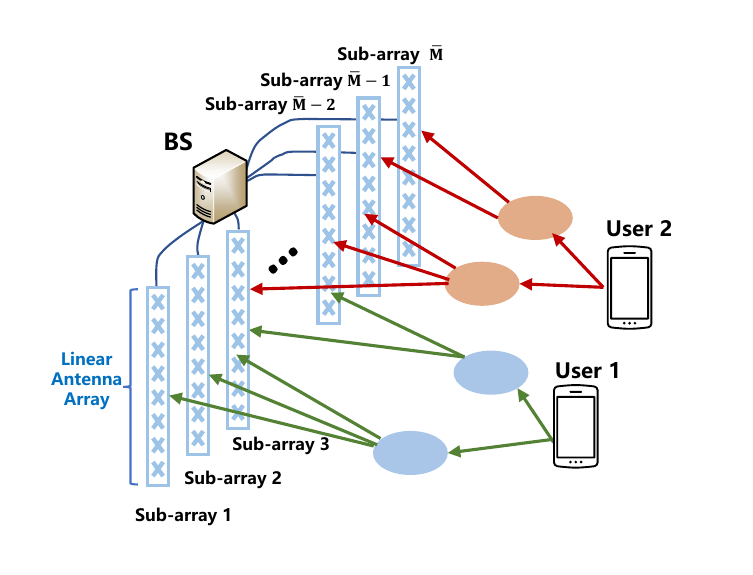}
\caption{Illustration of the multi-user XL-MIMO system with sub-array structured ELAA.  {The ELAA contains $\overline{M}$ antenna sub-arrays, with each sub-array being a linear antenna array.}  One can see that each scatter is visible by only a portion of sub-arrays -- demonstrating limited visibility region and hence spatial non-stationarity. In addition, far-away users are seen by different sub-arrays (or at different delay taps) because their active scatterers are mostly non-overlapped.}
\label{fig: XL-MIMO_system}
\end{figure}

\begin{figure*}[!t]
\centering
\begin{subfigure}[t]{0.45\textwidth}{
    \begin{minipage}{0.9\textwidth}
\includegraphics[width=\textwidth]{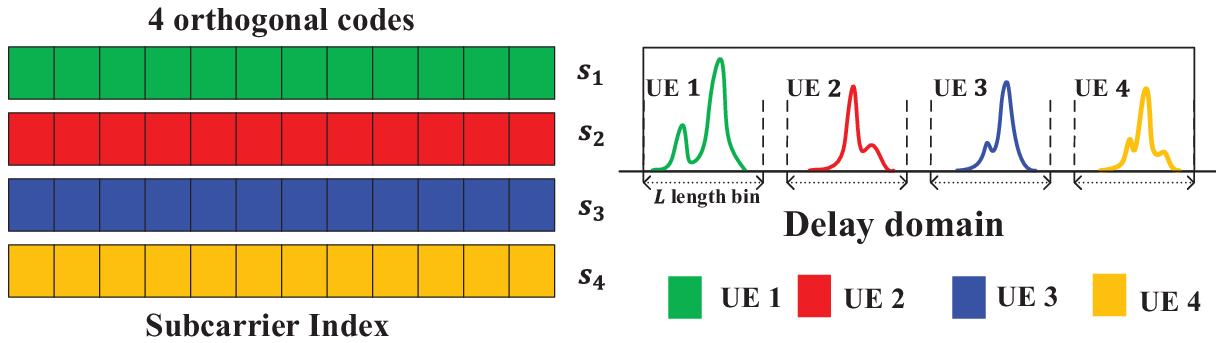}
	\caption{Orthogonal CDM}\label{fig:scheme-O-CDM}
    \end{minipage}
	}
\end{subfigure}
\begin{subfigure}[t]{0.45\textwidth}{
    \begin{minipage}{0.9\textwidth}
	\includegraphics[width=\textwidth]{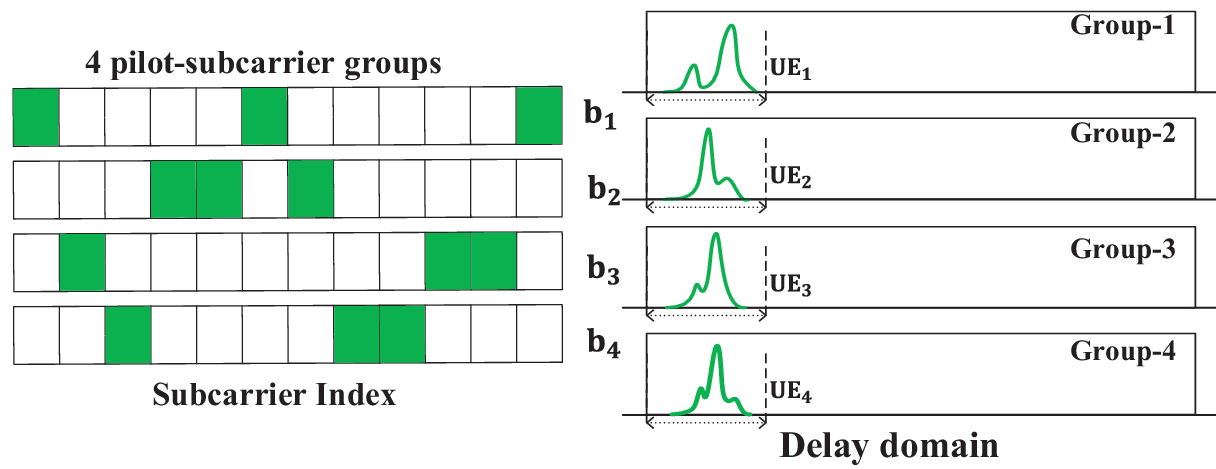}
	\caption{SR-FDM}\label{fig:scheme-SR-FDM}
    \end{minipage}
	}
\end{subfigure}
\begin{subfigure}[t]{0.45\textwidth}{
    \begin{minipage}{0.9\textwidth}
	\includegraphics[width=\textwidth]{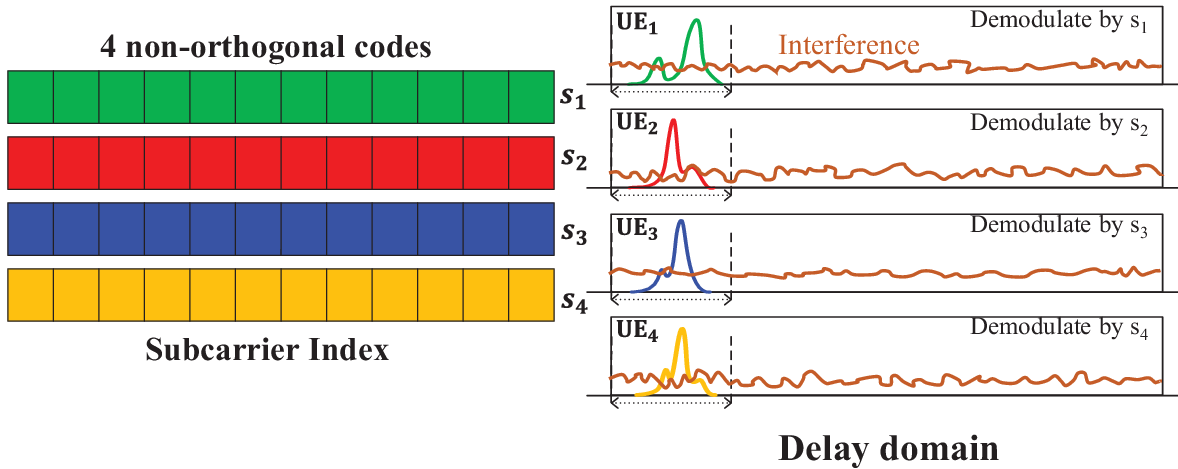}
	\caption{Non-orthogonal CDM}\label{fig:scheme-NO-CDM}
    \end{minipage}
	}
\end{subfigure}
\begin{subfigure}[t]{0.45\textwidth}{
    \begin{minipage}{0.9\textwidth}
	\includegraphics[width=\textwidth]{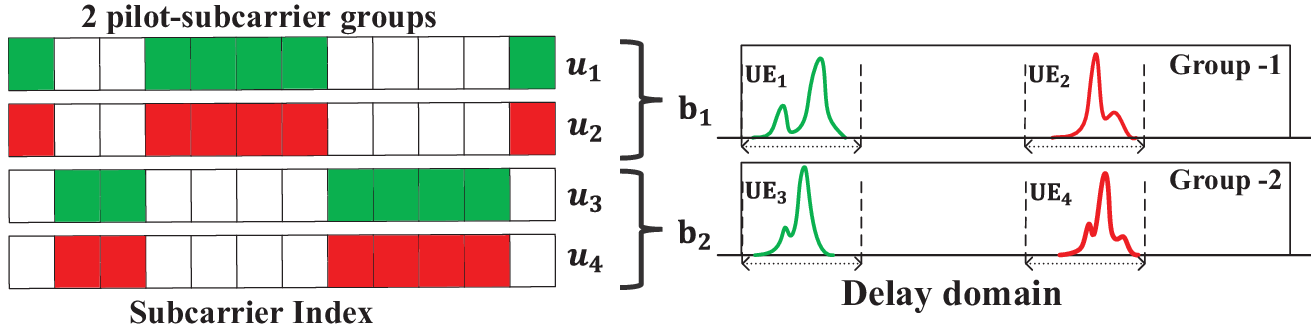}
	\caption{Proposed N-FD-CDM}\label{fig:scheme-N-FD-CDM}
    \end{minipage}
	}
\end{subfigure}
	\caption{Illustration of different pilot schemes for $K=4$.}
\label{fig:RefSigMechanism}
\end{figure*}

 {Pilot reduction schemes have been proposed for massive MIMO uplink channel sounding, where multiple users share one orthogonal frequency-division multiplexing (OFDM) pilot symbol. These schemes usually exploit the fact that the delay spread of channel impulse
response (CIR) is much smaller than the OFDM subcarrier length $N$.} Specifically, the orthogonal code-division multiplexing (O-CDM) pilot scheme, adopted by 5G New Radio (NR) systems \cite{B.Yang,H.Sahlin}  as shown in \figurename~\ref{fig:RefSigMechanism}(a), transmits cyclic shifted codes as pilot sequences. The  cyclic shifted codes adds $(k-1)L$ artificial delays to the CIR of the $k$-th user, and hence $\lfloor N/L\rfloor$ users can be estimated independently if $L$  exceeds the network’s maximum delay spread \cite{B.Yang,H.Sahlin}. This scheme does not fully exploit the delay-domain sparsity because $L$ is usually much larger than the CIR delay spread of an individual user \cite{S.Hou, X.Cheng}.  Based on this fact, sparse-recovery frequency-division multiplexing (SR-FDM) pilot scheme
 is proposed \cite{OAMP_MU, Rosati} to accommodate more users, as depicted in \figurename~\ref{fig:RefSigMechanism}(b).  However, all these schemes only exploit the delay-domain sparsity, which limits the number of supported users. In addition to these orthogonal pilot schemes, non-orthogonal CDM (NO-CDM) pilot scheme has been investigated in \cite{wang2014design},  where non-orthogonal sequences are transmitted by different users as illustrated in \figurename~\ref{fig:RefSigMechanism}(c). Although the number of supported users can be large with NO-CDM, the channel estimation suffers from severe inter-user interference. Therefore, a new pilot scheme is desirable for the XL-MIMO system to address the challenge posed by an increasing number of users.  

Channel estimation for ELAA is another critical issue, as the traditional sparse channel recovery algorithms suffer significant performance loss when users are located in the near-field region due to the large aperture of ELAA. Specifically, these algorithms promote angular-domain sparsity by leveraging the Discrete Fourier Transform (DFT) basis under the assumption that the spherical wavefronts can be well approximated by planer wavefronts \cite{channelmodel, CarlosF}. However, such an assumption does not hold in near-field cases. In addition, most existing XL-MIMO channel estimation literature focuses on such an issue by considering an ELAA with a large number of antennas spaced half a wavelength apart. In this case, the channel can be transformed into a sparse polar domain consisting of both angular and distance axes \cite{cui2022channel,lu2023near,lei2024hybrid,chen2021hybrid,yang2023practical}.  However, these studies require sparse-domain transformations over the entire ELAA, which heavily relies on the assumption of stationary channel statistics across all antennas. Unfortunately, such an assumption is invalid in XL-MIMO due to the limited visibility region effect, where scattering clusters may only be visible to a portion of antennas in the ELAA, leading to spatial non-stationarity \cite{yuan2022spatial}. Some recent works jointly address the limited visibility region and spherical wavefront issue for mmWave MIMO systems \cite{xu2024joint}. However, those methods often assume limited scattering clusters with small angular spreads, hence could not be applied in sub-6G systems. Additionally, the polar domain transform requires a half-wavelength spaced planar array setup, while the extremely larger number of antennas leads to significant infrastructure costs due to physical limitations \cite{yan2021joint,lu2014overview}.
 
In this paper, we investigate the multi-user uplink channel estimation in a sub-6G XL-MIMO OFDM system, in which the base station (BS) is equipped with a low-complexity ELAA composed of multiple sub-arrays. Each sub-array is a traditional MIMO array, and the distances between sub-arrays can be large and flexible to enhance the aperture size, as illustrated in Fig. \ref{fig: XL-MIMO_system}. This sub-array structured ELAA not only increases the illuminated area but also offers flexibility in its deployment in practice \cite{yan2021joint}. It also fully leverages traditional MIMO arrays and hence minimizes manufacturing variations for XL-MIMO systems. Due to the large sub-array separations, the limited visibility region effect becomes dominant.  Recent studies in \cite{iimori2022joint,yu2023channel,han2020channel} have proposed joint visibility detection and sub-array channel estimation algorithms for XL-MIMO systems. Further, in \cite{tang2024joint,tang2024spatially}, Markov prior is utilized to account for the high probability that adjacent sub-arrays are located in the visibility region of the same scattering cluster.   In  \cite{hou2019sparse,tan2023threshold,Chen2024}, channel estimation for XL-MIMO OFDM system is investigated by formulating a joint sparse channel recovery problem in both the antenna and delay domain. In \cite{cheng2019adaptive}, the structural sparsity in the antenna domain is exploited by utilizing a Dirichlet process.  In \cite{HMM}, structural sparsity features in antenna, delay, and user domain are jointly exploited by using multiple Markov prior models. However, the use of multiple one-dimensional Markov prior models is inefficient in capturing the clustering features across multiple domains. Moreover, all these algorithms are designed for orthogonal pilot schemes, which could not be directly applied to non-orthogonal pilot schemes.

To address all the above concerns, we propose a novel uplink non-orthogonal pilot scheme for a multi-user  XL-MIMO OFDM system with a sub-array structured ELAA. A customized channel estimation algorithm is proposed to suppress the inter-user interference by exploiting the structural channel sparsity in both antenna and delay domains. The main contributions of this paper are summarized as follows:

\begin{itemize}
	\item \textbf{Novel non-orthogonal-FD-CDM (N-FD-CDM) pilot sounding scheme for multi-user XL-MIMO}: We propose a novel N-FD-CDM pilot scheme for channel sounding in multi-user XL-MIMO systems with a sub-array structured ELAA. In this scheme, multiple users are grouped such that inter-group users utilize different pilot-subcarrier sets, while intra-group users are differentiated by distinct cyclic shifted codes, as illustrated in \figurename~\ref{fig:RefSigMechanism}(d). Unlike the conventional NO-CDM scheme, the proposed N-FD-CDM scheme effectively mitigates inter-group interference by leveraging the delay-domain sparsity. Compared to the SR-FDM scheme, the proposed N-FD-CDM scheme assigns multiple users to each pilot-subcarrier group, taking advantage of the fact that different users have distinct visibility regions in the ELAA, thereby significantly reducing the overall pilot overhead.
\item \textbf{Low-complexity Turbo-MRF algorithm based on approximate Bayesian inference with an MRF prior}:
     We employ a 2-dimensional (2-D) Markov Random Field (MRF) model to capture the individual and common clustered sparsity of users' CIRs in the antenna-delay domain,  {with much fewer hidden variables and hyper-parameters compared with \cite{HMM}.} Based on this, we propose a low-complexity Turbo-MRF algorithm for joint multi-user channel estimation, which consists of two modules: a group-wise linear minimum-mean-square-error (LMMSE) module with  {reduced complexity by frequency division of the N-FD-CDM pilot scheme}, and an MMSE module that infers the channel support of all users' CIRs using the 2-D MRF prior. Extensive simulations demonstrate that the proposed solution significantly outperforms state-of-the-art baselines while substantially reducing pilot overhead. 

\item \textbf{Low-complexity and effective pilot pattern optimization  {for frequency division}}:
 One issue associated with the proposed N-FD-CDM pilot scheme is the partial-DFT sensing matrix introduced by the frequency division between groups, which poses challenges for CIR recovery \cite{qi2012study,qi2014pilot}.  {To address this, we propose an effective pilot pattern optimization scheme by minimizing the mean squared error (MSE) of the initialized LMMSE estimator. The MSE objective is formulated to  account for inter-user interference and promote frequency division.} The resultant discrete optimization problem is relaxed into a continuous optimization problem by introducing a sparse-promoting reformulation. Simulations demonstrate that our optimized pilot pattern significantly improves robustness for joint multi-user channel estimation in XL-MIMO systems compared to randomly allocated pilot patterns. 

\end{itemize}

 \textit{Notations:}  
Throughout the paper,  $\mathrm{diag}(\mathbf{x})$($\mathrm{diag}(\mathbf{X})$) represents  {the diagonal matrix (vector) formed by $\mathbf{x}$ (matrix $\mathbf{X}$).}   $\mathbf{I}_{N_t}$ denotes $N_t\times N_t$ an identity matrix and $\mathbf{1}_N$ refers to all-one vector with length $N$. $(\cdot)^H$, $(\cdot)^*$ and $(\cdot)^{\mathrm{T}}$ represent the Hermitian, the conjugate and the transpose operations respectively.  Given a set $\kappa$, $|\kappa|$ refers to the cardinality of $\kappa$. $\int_{x_m^-}$ and $\sum_{\bf x}^{x_m^-}$ refer to integrating over $\bf x$ except $x_m$ and summing over  $\bf x$ except $x_m$, respectively. $x\propto y$ means $x$ is in proportion to $y$.

\section{System Model}\label{sec:System_model}
We consider a multi-user XL-MIMO OFDM system with one BS and multiple single-antenna users, as shown in {\figurename~\ref{fig: XL-MIMO_system}}. The BS  {is equipped} with a sub-array structured ELAA composed of $\overline{M}$ sub-arrays arranged horizontally, and each sub-array is a half-wavelength spaced uniform linear array (ULA) consisting of $\widetilde{M}$ antennas.
The total number of antennas is $M=\overline{M} \times \widetilde{M}$.
We focus on an uplink pilot sounding scheme, where each user transmits an $N$-point OFDM pilot symbol. We suppose that $K$ users share one OFDM pilot symbol and are divided into $G$ groups.

\subsection{Spatial Non-Stationarity of XL-MIMO Channel}
\label{sec:XL_MIMO_CH_Model}
Define the frequency domain channel response of the $k$-th ($k\in \{ 1,~\cdots,~K\}$) user as ${\bf H}_{k}=[\overline{\bf H}_{k,1},\overline{\bf H}_{k,2},\cdots,\overline{\bf H}_{k,\overline{M}}] \in {\mathbb C}^{N \times M}$. Herein, $\overline{\bf H}_{k,\overline{m}}=[\overline{\bf h}_{k,\overline{m},1},\overline{\bf h}_{k,\overline{m},2},\cdots,\overline{\bf h}_{k,\overline{m},N}]^{\rm T} \in {\mathbb C}^{N \times \widetilde{M}}, ~\forall \overline{m}  \in \big\{1,~\cdots,~\overline{M}\big\}$ and $\overline{\bf h}_{k,\overline{m},n}\in\mathbb {C}^{\widetilde{M}}, ~\forall n   \in \{1,~\cdots,~N\}$.   Given the sub-array structured ELAA, for the $\overline{m}$-th sub-array, the frequency response of user $k$ in the $n$-th subcarrier, denoted by {$\overline{\bf h}_{k,\overline{m},n} \in \mathbb {C}^{\widetilde{M}}$}, is modeled based on the planer wavefront assumption and is given by
\begin{equation}\label{equ:h_km}
\begin{aligned}[b]
\overline{\bf h}_{k,\overline{m},n}=  \sum_{r=1}^{R_{k}} v_{k,\overline{m},r}  {\beta_{k,\overline{m},r}} {\bf f}(\phi_{k,\overline{m},r})
e^{-{\jmath 2 \pi \frac{n}{N} \tau_{k,\overline{m},r}}}
,
\end{aligned}
\end{equation}
where  $R_{k}$ is the number of rays from user $k$ to  the $\overline{m}$-th sub-array. 
$v_{k,\overline{m},r}\in\{0,1\}$ is the visibility indicator of the $r$-th ray from user $k$ to the $\overline{m}$-th sub-array at the BS, representing the spatial non-stationarity that arises from the limited visibility region effect \cite{NonstationarityELAA,Flordelis,GaoX}. $\beta_{k,\overline{m},r}$, $\tau_{k,\overline{m},r}$ and $\phi_{k,\overline{m},r}$ are the complex-valued path degradation, path delay and the angle of arrival of the $r$-th ray at the $\overline{m}$-th sub-array, respectively. {${\bf f}(\phi_{k,\overline{m},r})\in \mathbb{C}^{\widetilde{M}}$} is the sub-array steering vector that is given by
\begin{equation}\label{equ:a_basis}
\begin{aligned}[b]
{\bf f}(\phi) = \frac{1}{\sqrt{\widetilde{M}}}
[1, e^{-\jmath \pi \cos\left(\phi\right) },\cdots,e^{-\jmath \pi (\widetilde{M}-1) \cos\left(\phi\right)}]^{\rm T}.
\end{aligned}
\end{equation}

In traditional massive MIMO systems, one may have the visibility indicator $v_{k,\overline{m},r}=1$ for all $\overline{m}$ and $\{\beta_{k,\overline{m},r},\tau_{k,\overline{m},r},\phi_{k,\overline{m},r}\}$ remains consistent across different $\overline{m}$. However, in XL-MIMO, both $v_{k,\overline{m},r}$ and $\{\beta_{k,\overline{m},r},\tau_{k,\overline{m},r},\phi_{k,\overline{m},r}\}$ vary between different sub-arrays due to the near-field spatial non-stationarity.
Consequently, naively adopting the channel estimation modeling and algorithm from the conventional MIMO systems can lead to suboptimal performance in XL-MIMO systems.

We can exploit the inherent channel sparsity in the antenna-delay domain by transforming the frequency channel response, $\overline{\bf H}_{k,\overline{m}}$, into the following expression
\begin{equation}\label{equ:H_gkm_to_X}
\begin{aligned}[b]
	\overline{{\bf H}}_{k,\overline{m}} = {\bf F}_{\rm{D},k} \overline{{\bf X}}_{k,\overline{m}}{\bf F}_{\widetilde{M}},
\end{aligned}
\end{equation}
where $\overline{{\bf X}}_{k,\overline{m}}\in\mathbb{C}^{L\times {\widetilde{M}}}$ is the antenna-delay domain CIR of user $k$ at the $\overline{m}$-th sub-array, and $L$ is larger than the network maximum delay spread.  {${\bf F}_{\rm{D},k}\in \mathbb{C}^{N\times L}$ is the partial $N$-DFT matrix that selects the $L$ columns corresponding to the delay domain CIR of user $k$. ${\bf F}_{\widetilde{M}}$ is the $\widetilde{M}$-DFT matrix, serving as the ULA basis according to \eqref{equ:a_basis}.

Collecting CIRs from all sub-arrays of user $k$ together, we have
\begin{equation}\label{equ:delay_to_freq_h}
\begin{aligned}[b]
	{\bf H}_{k} = {\bf F}_{\rm{D},k}{\bf X}_{k}{\bf F}_{\rm A},
\end{aligned}
\end{equation}
where  {${\bf X}_{k}=[\overline{{\bf X}}_{k,1},\overline{{\bf X}}_{k,2},\dots,\overline{{\bf X}}_{k,\overline{M}}] \in \mathbb{C}^{L\times M}$ and ${\bf F}_{\rm A}={\bf I}_{\overline{M}} \otimes {\bf F}_{\widetilde{M}} \in \mathbb{C}^{M\times M}$}.

Now, we elaborate the antenna-delay domain sparsity of $\overline{{\bf X}}_{k}$ by a toy example shown in  {\figurename~\ref{fig:XL_MIMO_channel_example}}. Therein, the CIRs are generated by the open source MATLAB toolbox of COST2100 model \cite{Cost_code} using scenario ``SemiUrban\_VLA\_2\_6GHz''. The supports of CIR \footnote{ {The supports of CIR refer to the set of CIR region whose value is non-zero (or the collection of the delay-antenna grids where valid paths arrive), and are marked in black in Fig. \ref{fig:XL_MIMO_channel_example}. Mathematically, for user $ k$, this is represented by the set $\{\overline{m},\ell | a_{k,\overline{m},\ell}\neq 0\} $ where $a_{k,\overline{m},\ell}$ is defined in \eqref{equ:prior_xkl} in Section \ref{sec: Bayesian Prior}.}} are decided by averaging the channel power across all antennas within a single sub-array. Two key features can be summarized from {\figurename~\ref{fig:XL_MIMO_channel_example}}:
\begin{itemize}
    \item \textbf{Distinct Sparsity Patterns Among Different Sub-Arrays and Users}: The channel supports exhibit sparse characteristics in the antenna-delay domain, with different sub-arrays displaying distinct sparsity patterns—referred to as spatial non-stationarity. Furthermore, these spatial sparsity patterns vary among different users and do not  {largely} overlap, which helps suppress inter-user interference and provides spatial orthogonality, thereby motivating our proposed N-FD-CDM pilot scheme.

   \item \textbf{2-D Clustered Sparsity for Sub-Array Channel Support}: The non-zero CIRs in the antenna-delay domain are 2-D clustered in distribution. This clustering occurs because adjacent sub-arrays are likely to be visible to the same scattering cluster (i.e., the same propagation ray) in the propagation environment, which promotes structured antenna-domain sparsity. Additionally, clustered scatterers usually arrive at sub-arrays within a short delay spread, which promotes structured delay-domain sparsity. The 2-D clustered distribution of CIRs can be leveraged in channel estimation algorithms for improved effectiveness and efficiency, as will be discussed in Section \ref{sec:Beyasian_model}.

\end{itemize}
\begin{figure}[!t]
	\centering
	\includegraphics[width=0.5\textwidth]{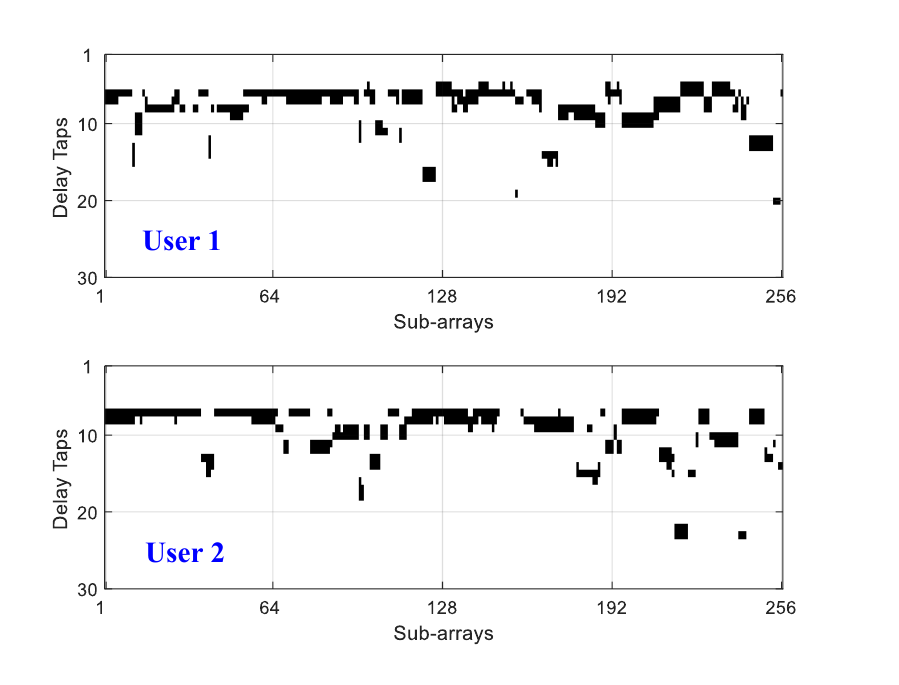}
	\caption{Supports of CIR realizations in antenna-delay domain for 512-OFDM symbol and 2 users with central frequency at 2.6 GHz. We set BS with 256 sub-arrays which are placed along $x$-axis with spacing 0.1 m, centered at (0, 0, 10)m. User-1 is located at (50, 70, 1.5)m, and User-2 is located at (-50, 100, 1.5)m.  {The figure shows that propagation paths from User 1 generally arrive at earlier delay taps (primarily between delay taps $5$ and $6$ along the y-axis) compared to those from User 2 (which are mostly between delay taps $7$ and $8$ along the y-axis).}}
	\label{fig:XL_MIMO_channel_example}
\end{figure}

\subsection{The Proposed N-FD-CDM Pilot Scheme}\label{sec:NFD_CDM_scheme}
The proposed N-FD-CDM pilot scheme is depicted in \figurename~\ref{fig:RefSigMechanism}(d) that integrates FDM and CDM. Specifically,
\begin{itemize}
    \item The FDM mechanism partitions the $N$ subcarriers into $G$ pilot-subcarrier groups, which can be represented by the binary selection vector $\overline{{\bf b}}_g =[\overline{b}_{g,1},\dots,\overline{b}_{g,N}]^{\mathrm{T}}   \in \{0,1\}^N,~\forall g\in \{1,~\cdots,~G\}$ with constraints
        \begin{align} 
        	\label{equ:pilot_con2}
        	\sum_{g = 1}^{G} \overline{b}_{g,n} &\leq 1, 
        \end{align}
        which guarantees the orthogonality among different groups.
    \item Within the same pilot-subcarrier group, the CDM mechanism distinguishes the intra-group users by assigning distinct cyclic shifted codes that introduce different artificial delays. Specifically, denote the user set in the $g$-th group by $\mathcal{K}_g$. The corresponding cyclic shifted codes for the $|\mathcal{K}_g|$ users in group $g$, denoted by  {$\widetilde{{\bf b}}_{g,k_g} =[\widetilde{b}_{g,k_g,1},\dots,\widetilde{b}_{g,k_g,N}]^{\text{T}} \in {\mathbb C}^{N},~\forall k_g\in\big\{1,~\cdots,~|\mathcal{K}_g|\big\}$} is given by
        \begin{align}
        	\label{equ:CDM}
        	\widetilde{b}_{g,k_g,n} = e^{-j2\pi\tau_{g,k_g}\left(n-1\right)/N},
        \end{align}
    which is the DFT transform of pulse sequence $\delta[t - \tau_{g,k_g}]$ for $t = \{1,~\dots,~N\}$, with $\tau_{g,k_g}$ being the artificial delay.
        For simplicity,
        let $\tau_{g,k_g} = \frac{{k_g}-1}{|\mathcal{K}_g|}{N}$ such that the CIRs of the users in group $g$ are uniformly spaced in the delay domain.
\end{itemize}

 {\textit{Remark 1:}  To compensate for the non-orthogonality among intra-group pilots,  we leverage the distinctly sparse and clustered CIR support among users during the channel estimation stage, as it will be detailed in Section \ref{sec:ALG}. Hence, our proposed N-FD-CDM pilot scheme yields a straightforward, practical, yet efficient grouping strategy: users are grouped according to a rule that maximizes the intra-group user distance \footnote{ {The group number $G$ can be determined based on the statistical sparsity of the environment, which tends to remain relatively stable over time. In this paper, we assume $K$ remains the same and set $G = K/2$ (or $|\mathcal{K}_g|=2$) without loss of generality. In the event of sudden user growth, we propose to maintain the sub-carrier group numbers for reasons of hardware efficiency, and assign new users to groups with users that are farther apart. However, we acknowledge that a more sophisticated grouping strategy could adapt better to sudden user scenarios, which presents an interesting avenue for future research.}}. }

Consequently, the pilot  {${\bf u}_{k} \in {\mathbb C}^{N}$} in the proposed N-FD-CDM scheme for the $k$-th user belonging to the $g$-th group is given by
\begin{align}\label{equ:NFD_CDM}
{\bf u}_{k} = \overline{{\bf b}}_g \odot \widetilde{{\bf b}}_{g,k_g},
\end{align}
where $\odot$ is the Hadamard product. Note that when $G = 1$, the N-FD-CDM scheme reduces to the traditional O-CDM scheme as shown in {\figurename~\ref{fig:RefSigMechanism}(a)}. When  {$G=K$ (or $|\mathcal{K}_g|=1, ~\forall g$)}, the N-FD-CDM scheme reduces to the traditional SR-FDM scheme as shown in {\figurename~\ref{fig:RefSigMechanism}(b)}.

The received OFDM signal ${\bf Y}\in {\mathbb C}^{N \times {M}}$ at the BS is then given by
\begin{equation}\label{equ:channel_vector1}
\begin{aligned}[b] 
{\bf Y}=\sum_{k=1}^K {\rm{diag}}({\bf u}_{k}){\bf H}_{k}+{\bf Z},
\end{aligned}
\end{equation}
where  ${\bf Z}\in {\mathbb C}^{N \times M} $ is the additive white Gaussian noise (AWGN), with independent and identically distributed (i.i.d.) elements following zero-mean complex Gaussian distribution ${\cal{CN}}(0,\sigma_{z}^2)$.

Thanks to the orthogonality among different pilot-subcarrier groups, the likelihood estimation for users' CIRs in different groups can be performed independently. Denote the received signal in the $g$-th pilot-subcarrier group by ${\bf Y}_{g} \in {\mathbb C}^{N \times M}$, which is expressed as follows
\begin{align}
	\label{equ:rx_sig_model}
	{\bf Y}_{g} &   = \sum_{k\in\mathcal{K}_g} \text{diag}({\bf u}_{k}){\bf H}_{k}+{\bf Z}_{g}\\
    \label{equ:rx_sig_model_2}
	 &= \sum_{k\in\mathcal{K}_g} \text{diag}({\bf u}_{k}){\bf F}_{\rm{D},k}{\bf X}_{k}{\bf F}_{\rm A}+{\bf Z}_{g}.
\end{align}
where  ${\bf Z}_g$ is the AWGN noise for subcarriers in group $g$.
The observation in \eqref{equ:channel_vector1} becomes ${\bf Y}=\sum_g {\bf Y}_{g}$.

For group $g$, we are only interested in frequency observations whose $\overline{b}_{g,n}=1$ -- referred to as  {effective} pilot subcarriers.  {Denoting by $B$ the length of effective pilots in ${{\bf u}}_{k}$, we extract the effective received signal $\widetilde{{\bf Y}}_{g}\in\mathbb{C}^{B\times M}$ from ${\bf Y}_{g}$, which is given by 
\begin{align}
	\label{equ:rx_sig_model_3}
	\widetilde{{\bf Y}}_{g} &= \sum_{k\in\mathcal{K}_g} \text{diag}(\widetilde{{\bf u}}_{k}) {\bf F}_{{{\rm P}},k}{\bf X}_{k}{\bf F}_{\rm A}+\widetilde{{\bf Z}}_{g}\in\mathbb{C}^{B\times M},
\end{align}
where $\widetilde{{\bf u}}_{k}\in \mathbb{C}^{B\times 1}$ is the  effective pilots extracted from ${{\bf u}}_{k}$. $\widetilde{{\bf Z}}_{g}\in\mathbb{C}^{B\times M}$} is the corresponding noise after extraction. ${\bf F}_{{\rm{P}},k}\in \mathbb{C}^{B\times L}$ is the partial DFT matrix which selects $B$ rows (corresponding to the effective pilot subcarrier indexes with $\overline{b}_{g,n}=1$ in group $g$) from $ {\bf F}_{\rm{D},k}$. 

 Consequently, the channel estimation for ${\bf H}_k$ is transformed into estimating the sparse CIR ${\bf X}_k$, which will be elaborated upon in Section \ref{sec:Beyasian_model} and Section \ref{sec:ALG}.

\section{Bayesian Modeling for CIRs in XL-MIMO}\label{sec:Beyasian_model}
 As summarized from Fig. \ref{fig:XL_MIMO_channel_example} in section \ref{sec:XL_MIMO_CH_Model}, in XL-MIMO systems, it is highly advantageous to propose a channel estimation solution that not only effectively addresses the spatial non-stationarity of CIRs but also exploits the inherent 2-D clustered sparsity in the antenna-delay domain for enhanced efficiency. To achieve this, we introduce a hierarchical Bayesian model with an MRF prior to characterize the CIR distributions.


\subsection{Group-wise Likelihood Functions}\label{sec:Bay Likelihood}
Write $\widetilde{{\bf Y}}_g=[\overline{{\bf Y}}_{g,1},\overline{{\bf Y}}_{g,2},\cdots,\overline{{\bf Y}}_{g,\overline{M}}]$
and $\widetilde{{\bf Z}}=[\overline{{\bf Z}}_{g,1},\overline{{\bf Z}}_{g,2},\cdots,\overline{{\bf Z}}_{g,\overline{M}}]$ where $\overline{{\bf Y}}_{g,\overline{m}}$ and $\overline{{\bf Z}}_{g,\overline{m}}$ denote the observations and noise at the $\overline{m}$-th sub-array in the $g$-th group, respectively. We have the following signal model
\begin{align}
	\label{equ:rx_sig_model_5}
	\overline{{\bf Y}}_{g,\overline{m}} &= \sum_{ k\in \mathcal{K}_g}  \text{diag}(\widetilde{{\bf u}}_{k}){\bf F}_{{\rm{P}},k} \overline{{\bf X}}_{k,\overline{m}}{\bf F}_{\widetilde{M}}+\overline{{\bf Z}}_{g,\overline{m}} \\\label{equ:rx_sig_model_6}
     &\triangleq \sum_{ k\in \mathcal{K}_g}  {\bf A}_k \overline{{\bf X}}_{k,\overline{m}}{\bf F}_{\widetilde{M}}+\overline{{\bf Z}}_{g,\overline{m}}\in\mathbb{C}^{B\times \widetilde{M}},    
\end{align}
where {${\bf A}_k=\text{diag}(\widetilde{{\bf u}}_{k}){\bf F}_{{\rm{P}},k}\in\mathbb{C}^{B\times L}$} is the sensing matrix for the $k$-th user.

Define $\overline{\mathcal{X}}_{g,\overline{m}}=\{\overline{{\bf X}}_{k,\overline{m}}:\forall k\in\mathcal{K}_g\}$ as the set of CIRs of users in group $g$ at the sub-array $\overline{m}$. From the signal model in \eqref{equ:rx_sig_model_6},  the likelihood function for $\overline{{\bf Y}}_{g,\overline{m}}$ prior on $\overline{\mathcal{X}}_{g,\overline{m}}$  is given by
\begin{equation}\label{equ:likelihood_subarray}
\begin{aligned}[b]
p\left(\overline{{\bf Y}}_{g,\overline{m}}|\overline{\mathcal{X}}_{g,\overline{m}}\right)
&=\frac{1}{(\pi \sigma_{z}^2)^{B{\widetilde{M}}}}
e^{-\frac{1}{\sigma_{z}^2} \left\| \overline{{\bf Y}}_{g,\overline{m}}- {\sum_{k\in \mathcal{K}_g}} {\bf A}_k \overline{{\bf X}}_{k,\overline{m}}{\bf F}_{\widetilde{M}} \right\|_{\rm F}^2  }.
\end{aligned}
\end{equation}

Then, define $\widetilde{\mathcal{X}}_g\in\Big\{\overline{\mathcal{X}}_{g,\overline{m}}:\forall \overline{m}=\{1,~\cdots,~\overline{M}\}\Big\}$ as the set of CIRs of users in group $g$. The overall likelihood function for $\widetilde{{\bf Y}}_g$ is given by
\begin{equation}\label{equ:likelihood}
\begin{aligned}[b]
p\left(\widetilde{{\bf Y}}_g|\widetilde{\mathcal{X}}_g\right)
&=\prod_{\overline{m}=1}^{\overline{M}} p\left(\overline{{\bf Y}}_{g,\overline{m}}|\overline{\mathcal{X}}_{g,\overline{m}}\right)\\
&=\frac{1}{(\pi \sigma_{z}^2)^{B{{M}}}}
e^{-\frac{1}{\sigma_{z}^2} \sum_{\overline{m}=1}^{\overline{M}} \left\| \overline{{\bf Y}}_{g,\overline{m}}- {\sum_{k\in \mathcal{K}_g}} {\bf A}_k \overline{{\bf X}}_{k,\overline{m}}{\bf F}_{\widetilde{M}} \right\|_{\rm F}^2  }.
\end{aligned}
\end{equation}

\subsection{Sparse Prior for CIRs}\label{sec: Bayesian Prior}
In the following, we build the prior distribution that captures the 2-D clustered sparsity of ${{\bf X}}_{k}$ in the antenna-delay domain. First, we introduce an aggregate channel support for each CIR sub-array. Specifically, we write $\overline{{\bf X}}_{k,\overline{m}}=[\overline{\bf x}_{k,\overline{m},1},\overline{\bf x}_{k,\overline{m},2},\cdots,\overline{\bf x}_{k,\overline{m},\ell}]^{\rm T}$ for {$\overline{\bf x}_{k,\overline{m},\ell}\in\mathbb{C}^{\widetilde{M}}$}  and define the channel support vector ${\bm \alpha}_{k,\overline{m}}=
[\alpha_{k,\overline{m},1},\alpha_{k,\overline{m},2},\cdots,\alpha_{k,\overline{m},L}]^{\rm T}\in\mathbb{C}^L$. The conditional prior distribution of   $\overline{{\bf X}}_{k,\overline{m}}$ is  given by
\begin{equation}\label{equ:prior_xkl}
\begin{aligned}[b]
p\left(\overline{{\bf X}}_{k,\overline{m}}|{\bm \alpha}_{k,\overline{m}}\right)
&=\prod_{\ell=1}^{L}p\left(\overline{\bf x}_{k,\overline{m},\ell}|\alpha_{k,\overline{m},\ell}\right)\\
&=\prod_{\ell=1}^{L}\prod_{m\in\mathcal{M}_{\overline{m}}} p\left({ x}_{k,m,\ell}|\alpha_{k,\overline{m},\ell}\right),
\end{aligned}
\end{equation}
where $\mathcal{M}_{\overline{m}}=\Big\{m:~\forall m\in \{(\overline{m}-1)\widetilde{M}+1,~\cdots,~\overline{m}\widetilde{M}\}\Big\}$ denotes the set of antennas that belong to sub-array $\overline{m}$.
$\alpha_{k,\overline{m},\ell}$ is the introduced channel support indicating
whether the CIR elements $\overline{ x}_{k,m,\ell},~\forall m\in\mathcal{M}_{\overline{m}},$ is zero (when $\alpha_{k,\overline{m},\ell}=0$) or one (when $\alpha_{k,\overline{m},\ell}=1$), and $p\left( { x}_{k,m,\ell}|\alpha_{k,\overline{m},\ell}\right)$ is given by
\begin{equation}\label{equ:prior_xkmn}
\begin{aligned}[b]
p\left({ x}_{k,m,\ell}|\alpha_{k,\overline{m},\ell}\right)
&=(1-\alpha_{k,\overline{m},\ell})\delta({ x}_{k,{m},\ell})+\\
&\alpha_{k,\overline{m},\ell} {\mathcal{CN}}({ x}_{k,m,\ell},0,\sigma^2_{k,\overline{m},\ell}),~m\in\mathcal{M}_{\overline{m}},
\end{aligned}
\end{equation}
in which ${\mathcal{CN}}({ x}_{k,m,\ell},0,\sigma^2_{k,\overline{m},\ell})$ represents that ${ x}_{k,m,\ell}$ follows an i.i.d complex Gaussian distribution with zero mean and variance $\sigma^2_{k,\overline{m},\ell}$ if non-zero.

From \eqref{equ:likelihood_subarray} to \eqref{equ:prior_xkmn},
it is sufficient to estimate the CIR $\overline{{\bf X}}_{k}$ from observations $\widetilde{Y}_g$ via Bayesian inference algorithms,
provided that the channel support ${\bm \alpha}_{k,\overline{m}}$ is known. Thus in the following, we build the prior distribution of ${\bm \alpha}_{k,\overline{m}}$ that specifically captures the inherent 2-D clustering features of CIRs.

\subsection{Hidden MRF Model for Channel Supports} \label{sec: Bayerian MRF}
In this subsection, we propose a hidden MRF prior model for the channel support ${\bm \alpha}_{k,\overline{m}}$, which exhibits clustered sparsity that could be leveraged for enhanced channel estimation. For this purpose, we introduce two aggregated latent variable vectors for  ${\bm \alpha}_{k,\overline{m}}$, denoted as  ${\bf s}_{k,\overline{m}}=[s_{k,\overline{m},1},s_{k,\overline{m},2},\cdots,s_{k,\overline{m},L}]^{\rm T}\in\mathbb{C}^L$ and 
${\bf s}_{c,\overline{m}}=[s_{{\rm c},\overline{m},1},s_{{\rm c},\overline{m},2},\cdots,s_{{\rm c},\overline{m},L}]^{\rm T}\in\mathbb{C}^L$. In this context, the conditional prior of the channel support vector ${\bm \alpha}_{k,\overline{m}}$ can be expressed as follows
\begin{equation}\label{equ:prior_alpha_km}
\begin{aligned}[b]
p\left({\bm \alpha}_{k,\overline{m}}|{\bf s}_{k,\overline{m}},{\bf s}_{c,\overline{m}}\right)
&=\prod_{\ell=1}^L p\left(\alpha_{k,\overline{m},\ell}|s_{k,\overline{m},\ell},{ s}_{c,\overline{m},\ell}\right),
\end{aligned}
\end{equation}
with
\begin{align} 
\notag&p\left(\alpha_{k,\overline{m},\ell}|s_{k,\overline{m},\ell},s_{{\rm c},\overline{m},\ell}\right)\\\notag
=&\left[1-\delta\left(s_{k,\overline{m},\ell}-1\right)\delta\left(s_{{\rm c},\overline{m},\ell}-1\right)\right]\delta\left(\alpha_{k,\overline{m},\ell}\right)+\\\label{equ:prior_alpha_kml}
		&\delta\left(s_{k,\overline{m},\ell}-1\right)\delta\left(s_{{\rm c},\overline{m},\ell}-1\right) \eta_k^{\alpha_{k,\overline{m},\ell}}(1-\eta_k)^{1-\alpha_{k,\overline{m},\ell}},
\end{align}
where $s_{k,\overline{m},\ell}\in\{-1,+1\}$ for $\forall \overline{m},\ell$ represents the 2-D MRF that indicates the  clustering properties of the individual CIR of user $k$.  {The individual 2-D MRF then allows for the incorporation of the clustered sparsity of CIR as the Bayesian prior for each user. Besides, users located in the same cell are likely to have their wireless signals reflected by the same scatterers before reaching the BS, and hence may exhibit similar clustering patterns. To leverage this prior knowledge, we introduce a common 2-D clustering support, denoted as $s_{{\rm c},\overline{m},\ell} \in \{-1, +1\}$, which represents the common clustered-scatterer pattern shared by all users within the same cell.} } \eqref{equ:prior_alpha_kml} means that $\alpha_{k,\overline{m},\ell}=1$ with probability $\eta_k$ only when $s_{k,\overline{m},\ell}=s_{{\rm c},\overline{m},\ell}=1$.

\begin{figure}[!t]
	\centering
	\includegraphics[width=0.3\textwidth]{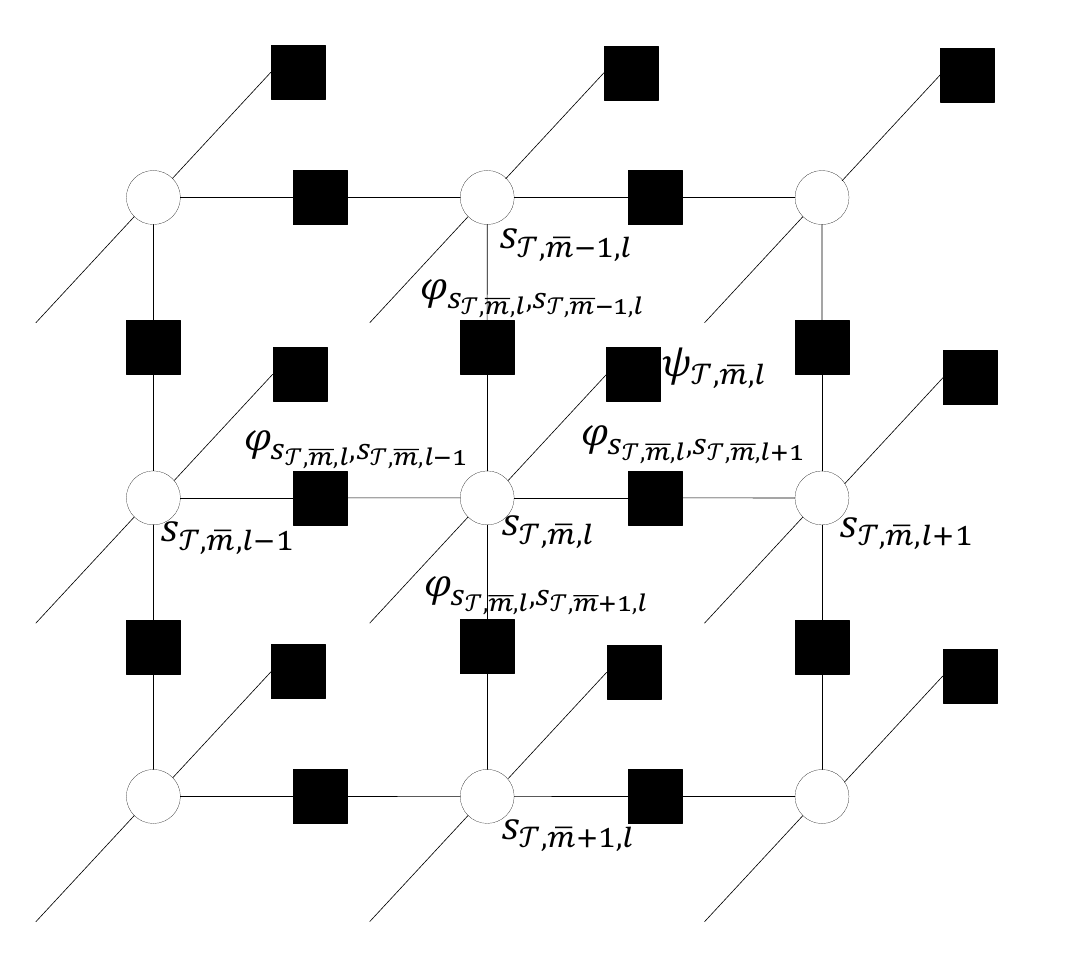}
	\caption{ {Illustration of the factor graph of MRF structure. In this figure, each hollow circle
represents a variable node $s_{\mathcal{T},\overline{m},\ell}$, while each black square represents a distribution function $\varphi(s_{\mathcal{T},\overline{m},\ell},s_{\mathcal{T},\overline{m'},\ell'})$ in \eqref{eq_MRF_func1} or $\psi(s_{\mathcal{T},\overline{m},\ell})$ in \eqref{eq_MRF_func2}.}}
	\label{fig:MRF}
\end{figure}

\begin{figure}[!t]
	\centering
	\includegraphics[width=0.4\textwidth]{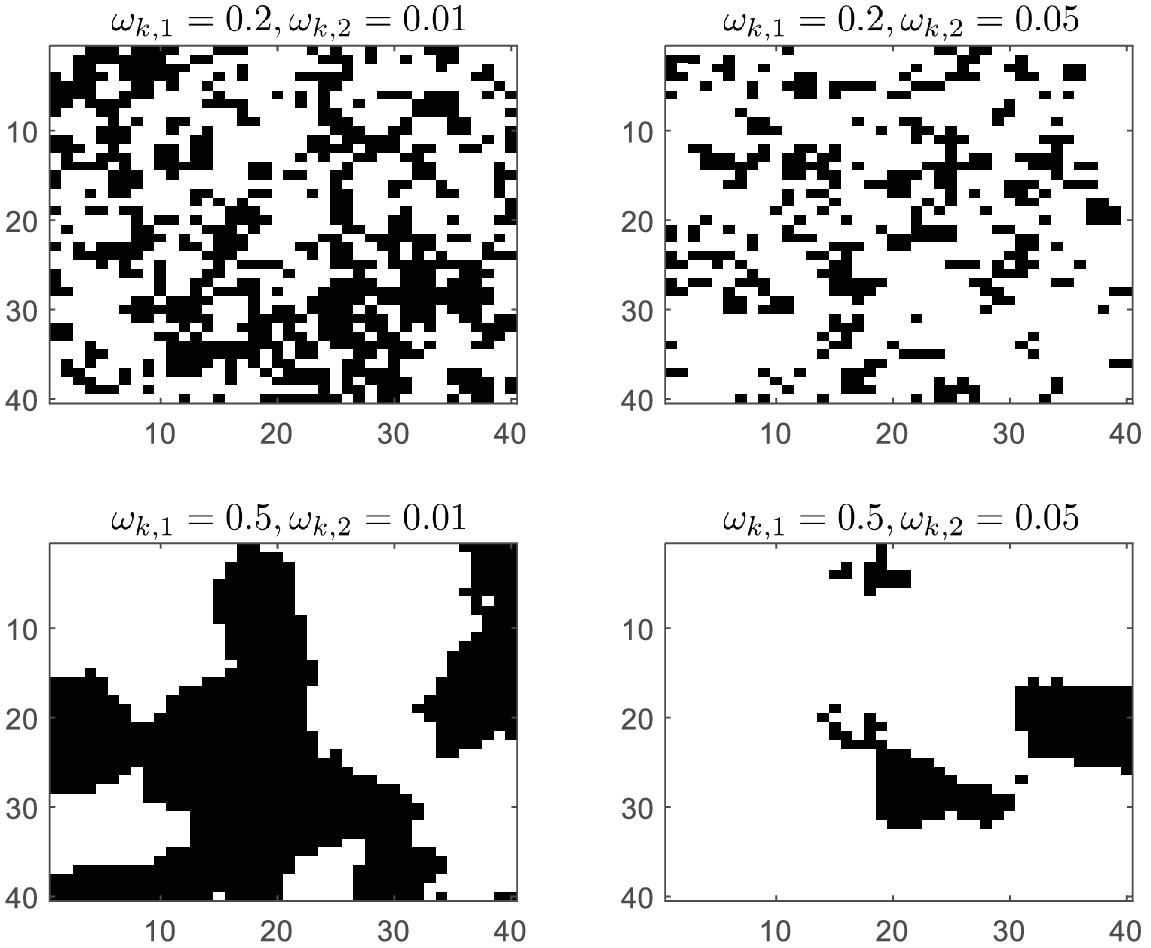}
	\caption{ {Random realizations of a $40\times40$ binary MRF ${\bf S}_{k}$ using distribution in \eqref{equ:MRF_Prob_1} with different MRF parameters. All the points satisfying ${s}_{k,\overline{m},\ell}=1$  are in black, and the rest are in white.}}
	\label{fig:MRF_example}
\end{figure}

The 2-D clustered prior is then captured  {by introducing} MRF structures to modeling $s_{k,\overline{m},\ell}$ and $s_{{\rm c},\overline{m},\ell}$. 
We first define ${\bf S}_{\mathcal{T}}=[{\bf s}_{\mathcal{T},1},{\bf s}_{\mathcal{T},2},\cdots,{\bf s}_{\mathcal{T},\overline{M}}] \in\{-1,1\}^{L\times \overline{M}},~\forall \mathcal{T}=\{{\rm c},1,~\cdots,~K\}$ as the aggregated latent variable matrix,  {where the subscript  $\mathrm{c}$ represents the common MRF pattern while the subscripts listing from $1$ to $K$ represent the private MRF pattern for user $k$.} This matrix features a binary MRF, as illustrated in Fig. \ref{fig:MRF}, where each random variable is connected to its 2-D neighbors.  {Mathematically}, we model ${\bf S}_\mathcal{T}$ by the 2-D Ising model with parameters $\mathcal{W}_\mathcal{T}=\{\omega_{\mathcal{T},1},\omega_{\mathcal{T},2}\}$ as follows
\begin{align}
	\label{equ:MRF_Prob_1}
	&p\left({\bf S}_{\mathcal{T}};\mathcal{W}_\mathcal{T}\right)\notag\\\propto&\underbrace{
\exp\left(\frac{1}{2}\sum_{\overline{m}=1}^{\overline{M}}\sum_{\ell=1}^{L}
\left(\sum_{\{\overline{m}',\ell'\}\in \mathcal{I}_{\overline{m},\ell} } \omega_{\mathcal{T},1}s_{\mathcal{T},\overline{m},\ell} s_{\mathcal{T},\overline{m}',\ell'}\right)
\right)
}_{\text{correlation term}}\notag\\
	&\times\underbrace{\exp\left(-\frac{1}{2}\sum_{\overline{m}=1}^{\overline{M}}\sum_{\ell=1}^{L} \omega_{\mathcal{T},2}s_{\mathcal{T},\overline{m},\ell} \right)}_{\text{local term}}\\
    =& \prod_{\overline{m},\ell}\psi(s_{\mathcal{T},\overline{m},\ell},s_{\mathcal{T},\overline{m'},\ell'})\left(\prod_{\{\overline{m}',\ell'\}\in \mathcal{I}_{\overline{m},\ell}}\varphi(s_{\mathcal{T},\overline{m},\ell},s_{\mathcal{T},\overline{m'},\ell'})\right),
\end{align}
where
\begin{align}\label{eq_MRF_func1}
\varphi(s_{\mathcal{T},\overline{m},\ell},s_{\mathcal{T},\overline{m'},\ell'})=&\exp\left(\frac{1}{2}\omega_{\mathcal{T},1}s_{\mathcal{T},\overline{m},\ell}s_{\mathcal{T},\overline{m'},\ell'}\right),\\\label{eq_MRF_func2}
    \psi(s_{\mathcal{T},\overline{m},\ell})=&\exp\left(-\frac{1}{2}\omega_{\mathcal{T},2}s_{\mathcal{T},\overline{m},\ell}\right),
\end{align}
where $\mathcal{I}_{\overline{m},\ell} = \{(\overline{m}-1,\ell),(\overline{m},\ell+1),(\overline{m}+1,\ell),(\overline{m},\ell-1)\}$.
In \eqref{equ:MRF_Prob_1}, the first 'correlation term' encourages $s_{\mathcal{T},\overline{m},\ell}$ to be identical to its neighbors, which promotes the clustered feature. The second 'local term' encourages $s_{\mathcal{T},\overline{m},\ell} = -1$, which promotes sparsity \cite{MRF_BOOK}.  {The effect of the MRF parameters is further demonstrated through examples in  {\figurename~\ref{fig:MRF_example}}, where increasing $\omega_{\mathcal{T},1}$ encourages clustering in ${\bf S}_{\mathcal{T}}$ and increasing $\omega_{\mathcal{T},2}$ enhances the sparsity of ${\bf S}_{\mathcal{T}}$.} 

\subsection{Overall Sparse Bayesian Model}\label{sec:Bayesian overall}


To sum up, the joint distribution of the observation $\widetilde{{\bf Y}}= \left[\widetilde{{\bf Y}}_1,~\cdots,~\widetilde{{\bf Y}}_G\right]$ and CIR $\widetilde{\mathcal{X}}=\{\widetilde{\mathcal{X}}_g:\forall g=1,~\cdots,~G\}$ given parameters $\widetilde{\mathcal{W}}=\{\widetilde{\mathcal{W}}_k:\forall k=1,~\cdots,~K\}$ is given
by
\begin{equation}\label{equ:overall_model}
\begin{aligned}[b]
p\left(\widetilde{{\bf Y}},\widetilde{\mathcal{X}};\widetilde{\mathcal{W}}\right)
&\propto\prod_{g}p\left(\widetilde{{\bf Y}}_g|\widetilde{\mathcal{X}}_{g}\right)\prod_{k}p\left({{\bf X}}_k;\mathcal{W}_k,\mathcal{W}_{\rm c}\right),
\end{aligned}
\end{equation}
where $p\left({{\bf X}}_{k};\mathcal{W}_k,\mathcal{W}_{\rm c}\right)$ is the overall prior probability of ${\bf {X}}_k$, which is modeled as follows
\begin{align}\label{equ:overall_model_pri}
p\left({{\bf X}}_k;\mathcal{W}_k,\mathcal{W}_{\rm c}\right)=&\sum_{\boldsymbol{\alpha}_k,{\bf S}_k,{\bf S}_{\rm c}}p({\bf S}_k;\mathcal{W}_k)p({\bf S}_{\rm c};\mathcal{W}_{\rm c})\times\notag\\
&\prod_{\overline{m}}p\left(\overline{{\bf X}}_{k,\overline{m}}|\boldsymbol{\alpha}_{k,\overline{m}}\right)p\left(\boldsymbol{\alpha}_{k,\overline{m}}|{\bf s}_{k,\overline{m}},{\bf s}_{{\rm c},\overline{m}}\right). 
\end{align}

\section{Low-complexity Inference Algorithm for multi-user XL-MIMO CE}\label{sec:ALG}

\begin{figure*}[t]
	\centering
	\includegraphics[width=0.85\textwidth]{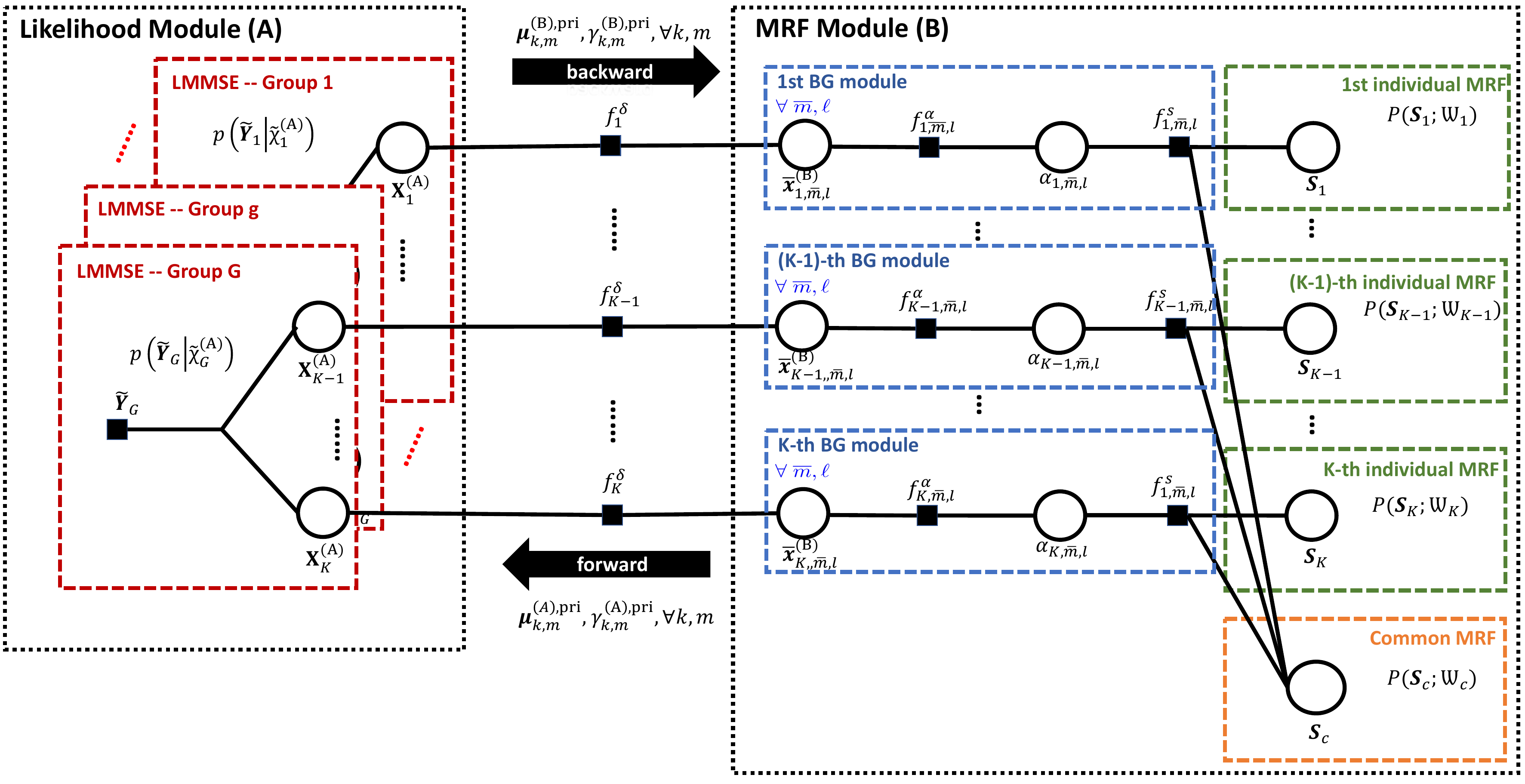}
	\caption{Factor graph of the hierarchical Bayesian model of the users' prior CIR in XL-MIMO.}
	\label{fig:factor_graph_overall}
\end{figure*}

Given the Bayesian model in \eqref{equ:overall_model}-\eqref{equ:overall_model_pri}, the  XL-MIMO channel estimation is formulated to determine the posterior mean of $x_{k,m,\ell}$ given observations $\widetilde{{\bf Y}}$, which is expressed as follows
\begin{equation}\label{equ:est_X}
\begin{aligned}
\hat{x}_{k,m,\ell}={\mathbb E} [{x}_{k,m,\ell}|\widetilde{\bf Y},\widetilde{\mathcal{W}}],
\end{aligned}
\end{equation}
whose expectation is over the following marginal posterior
\begin{equation}\label{equ:marginal_posterior}
\begin{aligned}[b]
p\left({x}_{k,m,\ell}|\widetilde{{\bf Y}},\widetilde{\mathcal{W}}\right)
&\propto \int_{ {x}_{k,m,\ell}^{-}}  p\left(\widetilde{{\bf Y}},\widetilde{\mathcal{X}};\widetilde{\mathcal{W}}\right).
\end{aligned}
\end{equation}

However, given the complexity of the Bayesian model specified in \eqref{equ:overall_model},  using brute-force marginalization for computing the posterior mean in \eqref{equ:est_X} is computationally prohibitive. To address this challenge, this section presents a low-complexity algorithm -- termed Turbo-MRF -- which leverages the Vector Approximate Message Passing (VAMP) algorithm to approximate the posterior inference of $\widetilde{{\bf X}}$ \cite{VAMP}. The factor graph of the VAMP algorithm and the involved Bayesian inference framework in our scheme are depicted in Fig. \ref{fig:factor_graph_overall}. All corresponding functions in  Fig. \ref{fig:factor_graph_overall} are listed in Table \ref{table:factorgraph_equation}. 

Specifically,  we define $\widetilde{\mathcal{X}}_g^{\rm (A)} = \{{\bf X}^{\rm (A)}_{k}: k \in \mathcal{K}_g\}$ as the CIRs of users in group $g$ updated by LMMSE in Module A of VAMP, where ${\bf X}^{\rm (A)}_{k} = [{\bf x}^{\rm (A)}_{k,1}, \dots, {\bf x}^{\rm (A)}_{k,M}]\in\mathbb{C}^{L\times M}$ and each ${\bf x}^{\rm (A)}_{k,m} = [{x}^{\rm (A)}_{k,m,1}, \dots, {x}^{\rm (A)}_{k,m,L}]\in\mathbb{C}^{L}$. Similarly, we define $\widetilde{\mathcal{X}}^{\rm (B)} = \{{\bf X}^{\rm (B)}_{k}: k =0,~\cdots,~K-1\}$ as the CIRs of the $k-th$ user in group $g$ updated by MMSE in Module B of VAMP, with ${\bf X}^{\rm (B)}_{k} = [{\bf x}^{\rm (B)}_{k,1}, \dots, {\bf x}^{\rm (B)}_{k,M}]$ and each ${\bf x}^{\rm (B)}_{k,m} = [{x}^{\rm (B)}_{k,m,1}, \dots, {x}^{\rm (B)}_{k,m,L}]$. Then, the joint distribution in \eqref{equ:overall_model} is reformulated as follows
\begin{align}
	\label{equ:factorized_PXY}
	&p\left(\widetilde{\bf Y},\widetilde{\mathcal{ X}};\widetilde{\mathcal{W}}\right)\notag\\
	=&\prod_{g}p\left(\widetilde{\bf Y}_{g}|\widetilde{\mathcal{ X}}^{\rm (A)}_{g}\right)\prod_{k}\delta\left({\bf X}_{k}^{\rm (A)}-{\bf X}_{k}^{\rm (B)}\right) \prod_{k}p\left({\bf X}_k^{\rm (B)};{\mathcal{W}_k},{\mathcal{W}_{\rm c}}\right),
\end{align}
which decouples the original joint distribution in \eqref{equ:overall_model} into two parts:  the likelihood probability of the observations conditioned on $\widetilde{\mathcal{X}}^{(\rm A)}_g$ in the first term, and the prior probability with respect to (w.r.t) $\widetilde{\mathcal{X}}^{(\rm B)}$ in the last term, as shown in Fig. \ref{fig:factor_graph_overall}. A delta function is used to link the CIR elements $\{{\bf X}_{k}^{\rm (A)}\}$ and $\{{\bf X}_{k}^{\rm (B)}\}$. In VAMP, posterior inference is computed in each module and transferred to the other module as prior knowledge for iterative updates.  If the VAMP algorithm converges, both posterior means of $\hat{
x}_{k,m,\ell}^{\rm (A)}$ and $\hat{ x}_{k,m,\ell}^{\rm (B)}$ will align with $ \mathbb{E}[{ x}_{k,m,\ell}|\widetilde{\bf Y}]$ in \eqref{equ:est_X}. 

In the following, we elaborate on each module's operation of VAMP in accordance with the rules of belief propagation.

\begin{table*}[!t]
	\renewcommand{\arraystretch}{1.3}
	\caption{\textcolor{black}{Factors Distribution For the Factor Graph in Fig. \ref{fig:factor_graph_overall}} }
	\label{table:factorgraph_equation}
	\centering
	\begin{tabular}{|c|c|c|}
		\hline
		Factor  & Distribution & \textcolor{black}{Function Expression}\\
		\hline
		$f^{\delta}$ & $p(x_{k,m,\ell}|\mu_{k,m,\ell},\gamma_{k,m,\ell})$& $f^{\delta}(x^{\rm (B)}_{k,m,\ell})=\mathcal{CN}\left(x^{\rm (B)}_{k,m,\ell};\mu^{\rm (B),pri}_{k,m,\ell},\gamma_{k,m}^{{\rm (B),pri}}\right)$\\
		\hline
		$f^{\alpha}$ & $p(x_{k,m,\ell}|\alpha_{k,\overline{m},\ell})$&$f^{\alpha}(x^{\rm (B)}_{k,m,\ell},\alpha_{k,\overline{m},\ell})=\left(1-\alpha_{k,\overline{m},\ell}\right)\delta\left(x^{\rm (B)}_{k,m,\ell}\right)+\alpha_{k,\overline{m},\ell}\mathcal{CN}\left(x^{\rm (B)}_{k,m,\ell};0,\sigma_{k,m,\ell}^2\right)$\\
		\hline
		$f^{s}$ & $p(\alpha_{k,\overline{m},\ell}|s_{k,\overline{m},\ell},s_{{\rm c},\overline{m},\ell})$&
\makecell{$ f^{s}(\alpha_{k,\overline{m},\ell},s_{k,\overline{m},\ell},s_{{\rm c},\overline{m},\ell})=\left[ 1-\delta\left(s_{k,\overline{m},\ell}-1\right)\delta\left(s_{{\rm c},\overline{m},\ell}-1\right) \right]\delta\left(\alpha_{k,\overline{m},\ell}\right)$\\$ \quad+\delta\left(s_{k,\overline{m},\ell}-1\right)\delta\left(s_{{\rm c},\overline{m},\ell}-1\right) \eta_k^{\alpha_{k,\overline{m},\ell}}(1-\eta_k)^{1-\alpha_{k,\overline{m},\ell}}$}\\
		\hline
		$\varphi_{{s_{\overline{m},\ell},s_{\overline{m}',\ell'}}}$ & $p(s_{\mathcal{T},\overline{m},\ell},s_{\mathcal{T},\overline{m}',\ell'})$ & \makecell{$\varphi({s_{\overline{m},\ell},s_{\overline{m}',\ell'}})=\exp\left(w_{\mathcal{T},1}s_{\mathcal{T},\overline{m},\ell}s_{\mathcal{T},\overline{m}',\ell'}/2\right),~\{\overline{m}',\ell'\}\in\mathcal{I}_{\overline{m},\ell}^{s}$}\\
				\hline
		$\psi_{\overline{m},\ell}$ & $p(s_{\mathcal{T},\overline{m},\ell})$ & $\psi({\overline{m},\ell})=\exp\left(-w_{\mathcal{T},2}s_{\mathcal{T},\overline{m},\ell}/2\right)$\\
             	\hline
	\end{tabular}
\end{table*}

\begin{table}[!t]  
	\centering
	\caption{ {Important Notations in Section \ref{sec:ALG}, where Prop. stands for properties, R.V. stands for random variable and Para. stands for parameters.} }
	\label{table:factorgraph_variable}
	\centering
	\begin{tabular}{|p{1.2cm}|p{0.5cm}|p{6.0cm}|}
		\hline
		Notations  & Prop. & {Explanation}\\
		\hline
		$x^{(A)/(B)}_{k,m,\ell}$   & R.V. & The CIR value for user $k$ at antenna $m$ and delay tap $\ell$ in Module A or Module B.\\\hline
        $\mathbf{x}^{\mathrm{(A),pri}}_{k,m}$ / $\mathbf{x}^{\mathrm{(B),pri}}_{k,m}$   & R.V. & The CIR vector, collecting all delay taps for user $k$ at antenna $m$, corresponding to the prior distribution input into  Module A or Module B.\\
        \hline
        $\mathbf{x}^{\mathrm{(A),pos}}_{k,m}$ / $\mathbf{x}^{\mathrm{(B),pos}}_{k,m}$   & R.V. & The CIR vector, collecting all delay taps for user $k$ at antenna $m$, corresponding to  the posterior distribution obtained after updating with the observations in Module A or the prior PDF in Module B.\\
		\hline
		$\alpha_{k,\overline{m},\ell}$    & R.V. & The CIR support that indicates whether a path arrives at sub-array $\overline{m}$ at delap tap $\ell$ for user $k$. \\
        \hline
		$s_{k,\overline{m},\ell}$ / $\mathbf{S}_{k}$   & R.V. & The aggregated latent variable captures the clustering property of $\alpha_{k,\overline{m},\ell}$ for user $k$, whose matrix $\mathbf{S}_k$ forming a joint Ising probability density function (PDF) as shown in \eqref{equ:MRF_Prob_1} over $ \overline{m}$ and $\ell$ \\
        \hline
		$s_{\mathrm{c},\overline{m},\ell} $ / $\mathbf{S}_{\mathrm{c}}$    & R.V. & The aggregated latent variable captures the shared clustering property among all users, whose matrix $\mathbf{S}_{\mathrm{c}}$ forming a joint Ising PDF in \eqref{equ:MRF_Prob_1} over $\overline{m}$ and $\ell$. \\
	\hline
$\boldsymbol{\mu}^{\mathrm{(A/B),pri}}_{k,m}$& Para. & The Gaussian mean for $\mathbf{x}^{\mathrm{(A/B),pri}}_{k,m}$. 
\\\hline $\boldsymbol{\mu}^{\mathrm{(A/B),pos}}_{k,m}$    & Para. & The Gaussian mean for  $\mathbf{x}^{\mathrm{(A/B),pos}}_{k,m}$. \\\hline
        $\gamma^{\mathrm{(A/B),pri}}_{k,m}$    & Para. & The Gaussian variance for $\mathbf{x}^{\mathrm{(A/B),pri}}_{k,m}$. \\
	\hline
  $\gamma^{\mathrm{(A/B),pos}}_{k,m}$    & Para. & The Gaussian variance for  $\mathbf{x}^{\mathrm{(A/B),pos}}_{k,m}$. \\
	\hline
	\end{tabular}
\end{table}

\subsection{Likelihood Estimation in Module A}
Following Fig. \ref{fig:factor_graph_overall}, Module A first performs forward propagation to infer the posterior distribution of $\mathbf{X}_k^{\rm (A)}$ based on the likelihood observations and the prior information from Module B. Subsequently, in the backward direction, it prepares the likelihood belief to be passed to Module B after extracting the relevant prior information.

\subsubsection{Forward propagation}
As shown in Fig. \ref{fig:factor_graph_overall}, Module B transfers the approximate belief of $\nu({\bf x}^{\rm (A),pri}_{k,m})=\mathcal{CN}({\bf x}^{\rm (A),pri}_{k,m};{\boldsymbol{\mu}}_{k,m}^{\rm (A),pri}, {\gamma}_{k,m}^{\rm (A),pri} {\bf I}_L)$ along the forward propagation. Combining with likelihood function in \eqref{equ:likelihood_subarray},  the posterior belief, $\nu({\mathbf{X}}_k^{\rm (A),pos})$, has the following form
\begin{align}
   \label{eqn:posterier_Module_A}
     \prod_{k} \nu({\mathbf{X}}_k^{\rm (A),pos})
     \propto&\prod_{g} p(\widetilde{\mathbf{Y}}_g|\widetilde{\mathcal{X}}_g^{\rm (A),pri})\prod_{k,m}\nu({\bf x}^{\rm (A),pri}_{k,m}).
\end{align}

Module A conducts group-wise belief propagation thanks to the orthogonality of pilot sub-carriers between groups. Referring to \cite{VAMP}, the posterior distribution of ${\bf x}^{\rm (A),pos}_{k,m}$ can be expressed as $\nu({\bf x}^{\rm (A),pos}_{k,m})\sim\mathcal{CN}({\bf x}^{\rm (A),pos}_{k,m};\hat{{\bf x}}_{k,m}^{\rm (A), pos}, \hat{\bf C}_{k,m}^{\rm (A),pos}$), whose $\hat{{\bf x}}_{k,m}^{\rm (A), pos}$ and $\hat{\bf C}_{k,m}^{\rm (A),pos}$ are given as follows
	\begin{align}
		\label{equ:module_A_post_mean}
		\hat{{\bf x}}_{k,m}^{\rm (A), pos}=& \gamma_{k,m}^{\rm (A),pri}{\bf A}_{k}^{\text{H}}\mathbf{M}_{g,m}\mathbf{m}_{g,m}+\boldsymbol{\mu}_{k,m}^{\rm (A),pri},\\
		\label{equ:module_A_post_var}
		\hat{\bf C}_{k,m}^{\rm (A),pos}=&\gamma_{k,m}^{\rm (A),pri}\mathbf{I}_L-\gamma_{k,m}^{\rm (A),pri,2}{\bf A}_{k}^{\text{H}}\mathbf{M}_{g,m}{\bf A}_{k},
	\end{align} 
    with 
    \begin{align}
        \label{eq_lmmse_matrix}\mathbf{M}_{g,m}=&\biggl(\sum_{k\in\mathcal{K}_g}\gamma_{k,m}^{\rm (A),pri}{\bf A}_{k}{\bf A}_{k}^{\text{H}}+\sigma_z^2{\bf I}_B\biggr)^{-1},\\
        \mathbf{m}_{g,m}=&\biggl({\bf y}_{g,m} - \sum_{k\in\mathcal{K}_g}{\bf A}_{k}\boldsymbol{\mu}_{k,m}^{\rm (A),pri}\biggr),
    \end{align}
    where we define $\widetilde{\mathbf{Y}}_g\mathbf{F}_{\mathrm{A}}^{\mathrm{H}}=\left[\mathbf{y}_{g,1},~\cdots,~\mathbf{y}_{g,M}\right]$ referring to \eqref{equ:delay_to_freq_h}.

\subsubsection{Backward propagation} 
In VAMP, the posterior covariance is averaged across tap delays for enhanced convergence. The averaged posterior variance to be transferred is represented as  $\widetilde{{\gamma}}_{k,m}^{\rm (A),pos}=\frac{1}{L}{\rm Tr}\left(\hat{\bf C}_{k,m}^{\rm (A),pos}\right)$. Consequently, the posterior message in Module A can be re-expressed as follows
\begin{align}
     \widetilde{\nu}({\mathbf{x}}_{k,m}^{\rm (A),pos})=\mathcal{CN}({{\bf x}}_{k,m}^{\rm (A), pos};\hat{{\bf x}}_{k,m}^{\rm (A), pos}, \widetilde{{\gamma}}_{k,m}^{\rm (A),pos} {\bf I}_L).
\end{align}

Then, the backward propagation in module A transfers its posterior likelihood to module B, denoted by $\nu({\bf x}^{\rm (B),pri}_{k,m})$, after decorrelating its input and output messages, i.e.,  $\nu({\bf x}^{\rm (B),pri}_{k,m})\propto  \widetilde{\nu}({\bf x}^{\rm (A),pos}_{k,m})/ \nu({\bf x}^{\rm (A),pri}_{k,m})$, which is given by
\begin{align}
    \nu({\bf x}^{\rm (B),pri}_{k,m})\sim \mathcal{CN}({\bf x}^{\rm (B),pri}_{k,m};\boldsymbol{\mu}_{k,m}^{\rm (B),pri}, \gamma^{\rm (B),pri}_{k,m}),
\end{align}
with
\begin{align}
	\label{equ:ext_A_mean}
	&\boldsymbol{\mu}_{k,m}^{\rm (B),pri}=\gamma^{\rm (B),pri}_{k,m}\left(\widetilde{{\gamma}}_{k,m}^{{\rm (A),pos},-1}\hat{{\bf x}}^{\rm (A)}_{k,m}-\gamma^{\rm (A),pri,-1}_{k,m}\boldsymbol{\mu}_{k,m}^{\rm (A),pri}\right),\\
	\label{equ:ext_A_var}
	&\gamma^{\rm (B),pri}_{k,m}=\left(\widetilde{{\gamma}}_{k,m}^{{\rm (A),pos},-1}-\gamma^{\rm (A),pri,-1}_{k,m}\right)^{-1}.
\end{align}

\subsection{Posterior Inference in Module B}\label{sec: ALG_B}
In Module B, the posterior inference of ${\bf X}_{k}^{\rm (B)}$ is calculated during backward propagation, by utilizing the  MRF prior and the prior likelihood message from Module A. Subsequently, during forward propagation, it prepares the approximate belief to be passed back to Module A.

\subsubsection{Backward propagation}
In Module B,  Module A gives the prior likelihood belief of 
$\nu({\bf x}_{k,m}^{\rm (B),pri})=\mathcal{CN}({\bf x}^{\rm (B),pri}_{k,m};{\boldsymbol{\mu}}_{k,m}^{\rm (B),pri}, {\gamma}_{k,m}^{\rm (B),pri}\mathbf{I}_L)$. Then, the posterior belief, $ \nu({\bf x}_{k,m}^{\rm (B),pos})$, can be inferred through a sum-product rule as follows 
\begin{align}
    \nu({\bf x}_{k,m}^{\rm (B),pos})\propto &\nu({\bf x}_{k,m}^{\rm (B),pri})\prod_{\ell}\nu_{f^{\alpha}\rightarrow x^{\rm (B)}_{k,m,\ell}}(x^{\rm (B)}_{k,m,\ell}),
\end{align}
where  $\nu_{f^{\alpha}\rightarrow x^{\rm (B)}_{k,m,\ell}}(x^{\rm (B)}_{k,m,\ell})$ is 
message passing  from  factor node $f^{\alpha}$ to $x^{\rm (B)}_{k,m,\ell}$ that carries the MRF prior information in \eqref{equ:overall_model_pri}.

 However, $\nu_{f^{\alpha}\rightarrow x^{\rm(B)}_{k,m,\ell}}(x^{\rm(B)}_{k,m,\ell})$ is computationally intractable due to the hierarchical prior model.  As a solution, we adopt message passing to approximate  $\nu_{f^{\alpha}\rightarrow x^{\rm(B)}_{k,m,\ell}}(x^{\rm(B)}_{k,m,\ell})$ following the factor graph in Fig. \ref{fig:factor_graph_overall} and the MRF sub-graph in Fig. \ref{fig:MRF}. Specifically, the message passing over the backward direction is described as follows.

\begin{itemize}
	\item \textbf{Message passing from $f^{\alpha}$ to $f^s$:} Along $f^{\alpha}\rightarrow \alpha_{k,\overline{m},\ell}\rightarrow f^{s}$, the message from factor node $f^{\alpha}$ to variable node $\alpha_{k,\overline{m},\ell}$ -- denoted as $\nu_{f^{\alpha}\rightarrow  \alpha_{k,\overline{m},\ell}}\left(\alpha_{k,\overline{m},\ell}\right)$ -- is given by
\begin{align}
	\label{equ:MSG_PASS_2}
	&\nu_{f^{\alpha}\rightarrow  \alpha_{ k,\overline{m},\ell}}\left(\alpha_{k,\overline{m},\ell}\right)\notag\\
 \propto&\int f^{\alpha}\left(x^{\rm (B)}_{k,m,\ell},\alpha_{k,\overline{m},\ell}\right)\nu_{x^{\rm (B)}_{k,m,\ell}\rightarrow f^{\alpha}}\left(x^{\rm (B)}_{k,m,\ell}\right)\text{d}x^{\rm (B)}_{k,m,\ell}\notag\\
        \propto& \pi_{n,m,k}^{{\rm 1},\text{in},\alpha}\delta\left(1-\alpha_{k,\overline{m},\ell}\right)+(1-\pi_{n,m,k}^{{\rm 1},\text{in},\alpha})\delta\left(\alpha_{k,\overline{m},\ell}\right),
\end{align}
where $\nu_{x^{\rm (B)}_{k,m,\ell}\rightarrow f^{\alpha}}\left(x^{\rm (B)}_{k,m,\ell}\right)=\nu_{f^\delta\rightarrow x^{\rm (B)}_{k,m,\ell}}(x^{\rm (B)}_{k,m,\ell})=\nu({\bf x}_{k,m}^{\rm (B),pri})$, which gives  $\pi_{k,m,\ell}^{{\rm 1},\text{in},\alpha}=\left(1+\frac{\mathcal{CN}\left(0;\mu^{\rm (B),pri}_{k,m,\ell},\gamma^{\rm (B),pri}_{k,m}\right)}{\mathcal{CN}\left(0;\mu^{\rm (B),pri}_{k,m,\ell},\gamma^{\rm (B),pri}_{k,m}+\sigma_{k,m,\ell}^2\right)}\right)^{-1}$. Then, the message from variable node $\alpha_{k,\overline{m},\ell}$ to factor node $f^s$ -- denoted as $\nu_{\alpha_{k,\overline{m},\ell}\rightarrow f^s}(\alpha_{k,\overline{m},\ell})$-- is given by
\begin{align}
	\label{equ:MSG_PASS_30}
	&\nu_{\alpha_{k,\overline{m},\ell}\rightarrow f^s}(\alpha_{k,\overline{m},\ell}) \notag\\
	\propto&\prod_{m\in\mathcal{M}_{\overline{m}}}\nu_{f^{\alpha}\rightarrow  \alpha_{k,\overline{m},\ell}}\left(\alpha_{k,\overline{m},\ell}=1\right) \delta(1-\alpha_{k,\overline{m},\ell})+\notag\\
	& \prod_{m\in\mathcal{M}_{\overline{m}}}\nu_{f^{\alpha}\rightarrow  \alpha_{k,\overline{m},\ell}}\left(\alpha_{k,\overline{m},\ell}=0\right) \delta(\alpha_{k,\overline{m},\ell})\notag\\
	\propto&\pi_{k,\overline{m},\ell}^{{\rm B},\text{out},\alpha}\delta\left(1-\alpha_{k,\overline{m},\ell}\right) + \left(1-\pi_{k,\overline{m},\ell}^{\rm B,\text{out},\alpha}\right)\delta\left(\alpha_{k,\overline{m},\ell}\right),
\end{align}
with $\pi_{k,\overline{m},\ell}^{{\rm 1},\text{out},\alpha}=\left(1+\frac{\prod_{m\in\mathcal{M}_{\overline{m}}}\left(1-\pi_{k,m,\ell}^{{\rm 1},\text{in},\alpha}\right)}{\prod_{m\in\mathcal{M}_{\overline{m}}}\pi_{k,m,\ell}^{{\rm 1},\text{in},\alpha}}\right)^{-1}$.

	\item \textbf{Message update in individual MRF:} Along $ f^s\rightarrow s_{k,\overline{m},\ell}\rightarrow f^s$, the message from factor node $f^s$ to variable node $s_{k,\overline{m},\ell}$ -- denoted as $\nu_{f^s\rightarrow s_{k,\overline{m},\ell}}\left(s_{k,\overline{m},\ell}\right)$ -- is given by
\begin{align}
	\label{equ:MSG_PASS_3}
	&\nu_{f^s\rightarrow s_{k,\overline{m},\ell}}\left(s_{k,\overline{m},\ell}\right) \notag\\
	=&\sum_{\alpha_{k,\overline{m},\ell},s_{{\rm c},\overline{m},\ell}}f^s\left(\alpha_{k,\overline{m},\ell},s_{k,\overline{m},\ell},s_{{\rm c},\overline{m},\ell}\right)\notag\\ &\nu_{\alpha_{k,\overline{m},\ell}\rightarrow f^s}\left(\alpha_{k,\overline{m},\ell}\right)\nu_{s_{{\rm c},\overline{m},\ell}\rightarrow f^s}\left(s_{{\rm c},\overline{m},\ell}\right)\notag\\
	\propto&\pi_{k,\overline{m},\ell}^{\text{in},s}\delta\left(1-s_{k,\overline{m},\ell}\right) + \left(1-\pi_{k,\overline{m},\ell}^{\text{in},s}\right)\delta\left(1+s_{k,\overline{m},\ell}\right),
\end{align}
with  $\pi_{k,\overline{m},\ell}^{\text{in},s}=\big(1+\frac{1-\pi_{k,\overline{m},\ell}^{{\rm 1},\text{out},\alpha}}{\left(1-\pi_{c,\overline{m},\ell}^{\text{out},s}\eta_k\right)\left(1-\pi_{k,\overline{m},\ell}^{{\rm 1},\text{out},\alpha}\right)+\pi_{c,\overline{m},\ell}^{\text{out},s}\pi_{k,\overline{m},\ell}^{{\rm 1},\text{out},\alpha}\eta_k} \big)^{-1}$, initialized by $\pi_{c,\overline{m},\ell}^{\rm {out},s}=0.5$. Sequentially, the individual MRF is updated according to Appendix \ref{sec:Appen MRF} with $I_{\rm mrf}$ iterations.  After the individual MRF update, the message fed back into factor node $f^s$ is given by
\begin{align}\label{equ:MSG_PASS_50}
\notag&\nu_{s_{k,\overline{m},\ell}\rightarrow f^s}(s_{k,\overline{m},\ell}) \\&\propto\pi_{k,\overline{m},\ell}^{\text{out},s}\delta\left(1-s_{k,\overline{m},\ell}\right) + \left(1-\pi_{k,\overline{m},\ell}^{\text{out},s}\right)\delta\left(1+s_{k,\overline{m},\ell}\right),
\end{align}
where $\pi_{k,\overline{m},\ell}^{\text{out},s}$ comes from \eqref{equ:MRF_msg_neighbor}-\eqref{equ:MRF_OutPut_Individual} in Appendix \ref{sec:Appen MRF}.

    \item \textbf{Message update in common MRF:} Along $f^s\rightarrow s_{{\rm c},\overline{m},\ell}\rightarrow f^s$, the message from factor node $f^s$ to variable node $s_{{\rm c},\overline{m},\ell}$ is given by
\begin{align}
	\label{equ:MSG_PASS_5}
	&\nu_{f^s\rightarrow s_{{\rm c},\overline{m},\ell}}\left(s_{{\rm c},\overline{m},\ell}\right)\notag\\
	&\propto\pi_{k,\overline{m},\ell}^{\text{in},s_{\rm c}}\delta\left(1-s_{{\rm c},\overline{m},\ell}\right) + \left(1-\pi_{k,\overline{m},\ell}^{\text{in},s_{\rm c}}\right)\delta\left(1+s_{{\rm c},\overline{m},\ell}\right),
\end{align}
with $\pi_{k,\overline{m},\ell}^{\text{in},s_{\rm c}}=\big(1+\frac{1-\pi_{k,\overline{m},\ell}^{{\rm 1},\text{out},\alpha}}{\left(1-\pi_{k,\overline{m},\ell}^{\text{out},s}\eta_k\right)\left(1-\pi_{k,\overline{m},\ell}^{{\rm 1},\text{out},\alpha}\right)+\pi_{k,\overline{m},\ell}^{\text{out},s}\pi_{k,\overline{m},\ell}^{{\rm 1},\text{out},\alpha}\eta_k} \big)^{-1}$. Collecting messages from all individual factor nodes, the message to the common MRF is given by $\pi_{c,\overline{m},\ell}^{\text{in},s}=\frac{\prod_k\pi_{k,\overline{m},\ell}^{\text{in},s_{\rm c}}}{\prod_k\pi_{k,\overline{m},\ell}^{\text{in},s_{\rm c}}+\prod_k\left(1-\pi_{k,\overline{m},\ell}^{\text{in},s_{\rm c}}\right)}$.
Then, the common MRF is updated by the iterative updating scheme given in Appendix \ref{sec:Appen MRF} with $I_{\rm mrf}$ iterations. After the common MRF update, the message fed back into factor node $f^s$ is given by
\begin{align}\label{equ:MSG_PASS_50_c}
\notag&\nu_{s_{{\rm c},\overline{m},\ell}\rightarrow f^s}(s_{{\rm c},\overline{m},\ell}) \\&\propto\pi_{k,\overline{m},\ell}^{\text{out},s_{{\rm c}}}\delta\left(1-s_{{\rm c},\overline{m},\ell}\right) + \left(1-\pi_{k,\overline{m},\ell}^{\text{out},s_{\rm c}}\right)\delta\left(1+s_{{\rm c},\overline{m},\ell}\right),
\end{align}
where $\pi_{k,\overline{m},\ell}^{\text{out},s_{\rm c}}$ comes from \eqref{equ:MRF_msg_neighbor}-\eqref{equ:MRF_OutPut_common} in Appendix \ref{sec:Appen MRF}.

\end{itemize}

\subsubsection{Forward propagation}
In forward propagation, we conduct message passing from $f^s$ to $f^{\alpha}$ to finally obtain $\nu_{f^{\alpha}\rightarrow x^{\rm(B)}_{k,m,\ell}}(x^{\rm(B)}_{k,m,\ell})$. Specifically, along  $f^s\rightarrow\alpha_{k,\overline{m},\ell}\rightarrow f^\alpha\rightarrow x^{\rm(B)}_{k,m,\ell}$,  the message from factor node $f^s$ to variable node $\alpha_{k,\overline{m},\ell}$ is given by
\begin{align}
	\label{equ:MSG_PASS_8}
	&v_{f^s\rightarrow \alpha_{k,\overline{m},\ell}}\left(\alpha_{k,\overline{m},\ell}\right)\notag\\
	\propto&\pi_{k,\overline{m},\ell}^{{\rm 2,in},\alpha}\delta\left(1-\alpha_{k,\overline{m},\ell}\right) + \left(1-\pi_{k,\overline{m},\ell}^{{\rm 2,in},\alpha}\right)\delta\left(1+\alpha_{k,\overline{m},\ell}\right),
\end{align}
where $\pi_{k,\overline{m},\ell}^{{\rm 2,in},\alpha}=\pi_{k,\overline{m},\ell}^{\text{out},s}\pi_{k,\overline{m},\ell}^{\text{out},s_{\rm c}}\eta_k$.  Then, the message from factor node $f^\alpha$ to variable node $x_{k,m,\ell}$ is given by
\begin{align}
	\label{equ:MSG_PASS_9}
	\nu_{f^\alpha\rightarrow x^{\rm (B)}_{k,m,\ell}}\left(x^{\rm (B)}_{k,m,\ell}\right)\propto&\left(1-\pi_{k,m,\ell}^{\rm 2,\text{out},\alpha}\right)\delta\left(x^{\rm (B)}_{k,m,\ell}\right)+ \notag\\
	& \pi_{k,m,\ell}^{\rm 2,\text{out},\alpha}\mathcal{CN}\left(x^{\rm (B)}_{k,m,\ell};0,\sigma_{k,m,\ell}^2\right),
\end{align}
with $\pi_{k,m,\ell}^{\rm 2,\text{out},\alpha}=\left( 1+\frac{(1-\pi_{k,\overline{m},\ell}^{\rm 2,in,\alpha})\prod_{m'\in\mathcal{M}_{\overline{m}}}^{m^-}(1-\pi_{k,m,\ell}^{\rm 1,\text{in},\alpha})}{\pi_{k,\overline{m},\ell}^{\rm 2,in,\alpha}\prod_{m'\in\mathcal{M}_{\overline{m}}}^{m^-}\pi_{k,m,\ell}^{\rm 1,\text{in},\alpha}}\right)^{-1}$.

In the end, the posterior inference at variable node $x^{\rm (B)}_{k,m,\ell}$ is given by  
\begin{align}
    \label{eqn:posterior_b}
\notag\nu({x}_{k,m,\ell}^{\rm (B),pos})\propto &\nu({x}_{k,m,\ell}^{\rm (B),pri})\nu_{f^{\alpha}\rightarrow x^{\rm (B)}_{k,m,\ell}}(x^{\rm (B)}_{k,m,\ell})\\
\propto &\mathcal{CN}({x}_{k,m,\ell}^{\rm (B),pos};\hat{x}^{\rm (B),pos}_{k,m,\ell},\hat{\gamma}^{\rm (B),pos}_{k,m,\ell}),
\end{align}
with
\begin{align}
	\label{equ:module_B_post_mean}
	\hat{x}^{\rm (B),pos}_{k,m,\ell}=&\int x^{\rm (B),pos}_{k,m,\ell}\nu({x}_{k,m,\ell}^{\rm (B),pos})={\pi}_{k,m,\ell}\hat{r}_{k,m,\ell},\\
	\label{equ:module_B_post_var}
	\hat{\gamma}^{\rm (B),pos}_{k,m,\ell}
	 =&\int\left|x^{\rm (B),pos}_{k,m,\ell} - \mathbb{E}\left[x^{\rm (B),pos}_{k,m,\ell}\right]\right|^2\nu({x}_{k,m,\ell}^{\rm (B),pos})\\\notag
     =&{\pi}_{k,m,\ell}(\hat{C}_{k,m,\ell}+|\hat{r}_{k,m,\ell}|^2)-\pi^{2}_{k,m,\ell}|\hat{r}_{k,m,\ell}|^2, 
     \end{align}
where ${\pi}_{k,m,\ell}=\left[1+\frac{1-\pi^{\rm R, out, \alpha}_{k,m,\ell}}{\pi^{\rm R, out, \alpha}_{k,m,\ell}}\frac{\mathcal{CN}(\mu^{\rm (B),pri}_{k,m,\ell};0,\gamma_{k,m}^{\rm (B),pri})}{\mathcal{CN}(\mu^{\rm (B),pri}_{k,m,\ell};0,\sigma^2_{k,\overline{m},\ell}+\gamma_{k,m}^{\rm (B),pri})}\right]^{-1}$ is the posterior probability of ${ x}^{\rm(B)}_{k,m,\ell}$ being non-zero, $
     \hat{r}_{k,m,\ell}={\mu^{\rm (B),pri}_{k,m,\ell}\gamma^{{\rm (B),pri},-1}_{k,m}}/\big[{\gamma_{k,m}^{{\rm (B),pri},-1}+\sigma^{-2}_{k,\overline{m},\ell}}\big]$, and $
     \hat{C}_{k,m,\ell}= \big[\gamma_{k,m}^{{\rm (B),pri},-1}+\sigma^{-2}_{k,\overline{m},\ell}\big]^{-1}$ \cite{vila2011expectation}.

 Then, we average the posterior covariance over the tap delays similarly to Module A, which is represented as  $\widetilde{\gamma}_{k,m}^{\rm (B),pos}=\frac{1}{L}\sum_{\ell}\hat{  \gamma}_{k,m,\ell}^{\rm (B),pos}$. Hence, denoting by $\hat{\mathbf{x}}_{k,m}^{\rm (B),pos}=[\hat{ {x}}_{k,m,1}^{\rm (B),pos},~\cdots,~\hat{ {x}}_{k,m,L}^{\rm (B),pos}]^{\mathrm{T}}$, the posterior message in Module B can be approximated as follows
\begin{align}
     \widetilde{\nu}({\mathbf{x}}_{k,m}^{\rm (B),pos})=\mathcal{CN}({{\bf x}}_{k,m}^{\rm (B), pos};\hat{{\bf x}}_{k,m}^{\rm (B), pos}, \widetilde{\gamma}_{k,m}^{\rm (B),pos} {\bf I}_L).
\end{align}
Then, the backward propagation in module B transfers its  MRF prior to module A, denoted by $\nu({\bf x}^{\rm (A),pri}_{k,m})$, after decorrelating its input and output messages, i.e.,  $\nu({\bf x}^{\rm (A),pri}_{k,m})\propto  \widetilde{\nu}({\bf x}^{\rm (B),pos}_{k,m})/ \nu({\bf x}^{\rm (B),pri}_{k,m})$, which is given by  \cite{VAMP} 
\begin{align}
    \nu({\bf x}^{\rm (A),pri}_{k,m})=\mathcal{CN}({\bf x}^{\rm (A),pri}_{k,m};\boldsymbol{\mu}_{k,m}^{\rm (A),pri}, \gamma^{\rm (A),pri}_{k,m} {\bf I}_L),
\end{align}
with
\begin{align}
	\label{equ:ext_B_mean}
	&\boldsymbol{\mu}_{k,m}^{\rm (A),pri}=\gamma^{\rm (A),pri}_{k,m}\left(\widetilde{\gamma}^{{\rm (B),pos}-1}_{k,m}\hat{{\bf x}}^{\mathrm{(B),pos}}_{k,m}-\gamma^{\rm (B),pri,-1}_{k,m}\boldsymbol{\mu}_{k,m}^{\rm (B),pri}\right),\\
	\label{equ:ext_B_var}
	&\gamma^{\rm (A),pri}_{k,m}=\left(\widetilde{\gamma}^{{\rm (B),pos}-1}_{k,m}-\gamma^{\rm (B),pri,-1}_{k,m}\right)^{-1}.
\end{align}\par

\subsection{Complexity analysis}  { The proposed Turbo-MRF algorithm iteratively computes posterior inference for Module A and Module B until convergence is achieved. In Module A, the primary source of complexity arises from calculating $\mathbf{M}_{g,m}$ in the  LMMSE estimator which requires matrix inversion, as defined in \eqref{eq_lmmse_matrix}. In this context, the proposed N-FD-CDM pilot scheme enables a group-wise LMMSE estimator that features reduced dimensions through the frequency division strategy, resulting in a computational complexity of $\mathcal{O}(B^2(LK/G + B))$ per calculation. This is significantly lower than the complexity of the traditional CDM pilot scheme, which is $\mathcal{O}(N^3K+N^3)$ (where $N \gg B$)  per calculation. Additionally, this complexity can be further reduced by employing Schur Complement Computation for the matrix inverse in the special case of $K/G = 2$.}

 {In Module B, the proposed Turbo-MRF algorithm also benefits from reduced computational complexity, due to the structured ELAA. Specifically, the structured ELAA addresses the near-field and spatial non-stationarity effects at the level of the antenna sub-array rather than at each individual antenna. Consequently, it allows reduced  overall computational complexity (originating from reducing the number of variables involved in the factor graph of the Bayesian prior) in Module B from $\mathcal{O}(MKL)$ to $\mathcal{O}(\overline{M}KL)$ (where $M\gg \overline{M}$). Additionally, by incorporating the MRF prior discussed in Section \ref{sec: Bayesian Prior}, the number of hyper-parameters in the Bayesian prior is significantly reduced \cite{HMM}, facilitating low complexity and fast convergence.}

\subsection{Parameter Updates}\label{sec:ALG_EM}
In the proposed algorithm, the fitness of unknown parameters in  {module B -- denoted} by $\mathbf{q}=\big[\{\eta_k\},\{\sigma^2_{k,\overline{m},\ell}\},\{\mathcal{W}_k\},\{\mathcal{W}_{\rm c}\} \big]$ -- affects the performance and convergence. In this sub-section, we update the parameters following an EM rule \cite{vila2011expectation}. The EM algorithm updates each parameter iteratively by maximizing the posterior expectation of the log-likelihood function of the prior distribution in \eqref{equ:overall_model_pri}. For low complexity, we update each individual parameter while fixing the others iteratively. Consequently, at the $(i+1)$-th iteration, the parameters are updated by (Note that $\{\mathcal{W}_k\}$ and $\{\mathcal{W}_{\rm c}\} $ could be updated according to \cite[Result 1.1]{ghosal2020joint} using gradient descent.)
     \begin{align}
        \label{equ:update_eta_EM}
        \eta^{i+1}_k=&\frac{\sum_{\overline{m},\ell}p(\alpha_{k,\overline{m},\ell}=1\big|\mathbf{q}^{i})}{\sum_{\overline{m},\ell}p(s_{k,\overline{m},\ell}=1\big|\mathbf{q}^{i})p(s_{{\rm c},\overline{m},\ell}=1\big|\mathbf{q}^{i})},\\
        \label{equ:update_sigma_EM}\sigma^{2,i+1}_{k,\overline{m},\ell}=&\frac{\sum_{m\in\mathcal{M}_{\overline{m}}}{\pi}_{k,m,\ell}(|\hat{r}_{k,m,\ell}|^2+ \hat{C}_{k,m,\ell})}{\sum_{m\in\mathcal{M}_{\overline{m}}}p(\alpha_{k,\overline{m},\ell}=1\big|\mathbf{q}^{i})},
    \end{align}
    with
    \begin{align}
        \notag p(\alpha_{k,\overline{m},\ell}\big|\mathbf{q}^{i})\propto& \nu_{\alpha_{k,\overline{m},\ell}\rightarrow f^s}(\alpha_{k,\overline{m},\ell})\nu_{ f^{\alpha}\rightarrow\alpha_{k,\overline{m},\ell}}(\alpha_{k,\overline{m},\ell}),\\
        \notag p(s_{k,\overline{m},\ell}\big|\mathbf{q}^{i})\propto& \nu_{ f^{s}\rightarrow s_{k,\overline{m},\ell}}(s_{k,\overline{m},\ell})\nu_{s_{k,\overline{m},\ell}\rightarrow f^s}(s_{k,\overline{m},\ell}).
    \end{align}

 {More details about the EM updates are uncovered in Appendix B. }We summarize the  Turbo-MRF algorithm in Algorithm \ref{ALG:alg1}.
\begin{algorithm}[!tbh]
	\caption{ Turbo-MRF Algorithm}
	\label{ALG:alg1}
	\begin{algorithmic}[1]
		\renewcommand{\algorithmicrequire}{\textbf{Input:}}
		\renewcommand{\algorithmicensure}{\textbf{Output:}}
		\REQUIRE Received signal $\{{\bf Y}_1,\dots,{\bf Y}_G\}$, sensing matrix $\{{\bf A}_1,\dots,{\bf A}_K\}$, AWGN noise covariance ${\bf C}_{{\bf z}}$, maximum iteration number $I_{\text{max}}$, MRF inner iteration number $I_{\text{mrf}}$.
		\ENSURE  Estimated channel matrix $\{\hat{{\bf H}}_1,\dots,\hat{{\bf H}}_K\}$.
		\\ \textbf{Initialization:} $\boldsymbol{\mu}_{k,m}^{\rm (A),pri} = {\bf 0}$, $\gamma_{k,m}^{\rm (A), pri} = \frac{N}{L}$.
		\FOR{$iter \leq I_{\text{max}}$}
		\STATE \textbf{\%Module A:} \\
		\STATE Update the posterior mean and variance of ${\bf x}_{k,m}^{\rm (A)}$: $\hat{{\bf x}}_{k,m}^{\rm (A),pos}$, $\hat{\mathbf{C}}_{k,m}^{\rm (A),pos}$ with (\ref{equ:module_A_post_mean}) - (\ref{equ:module_A_post_var}).\\
		\STATE Update the extrinsic messages $\boldsymbol{\mu}_{k,m}^{\rm (B),pri}$ and $\gamma_{k,m}^{\rm (B),pri}$ with (\ref{equ:ext_A_mean}) - (\ref{equ:ext_A_var}).
		\STATE \textbf{\%Module B :}\\
		\STATE Message passing over $f^\delta\rightarrow x^{\rm (B)}_{k,m,\ell}\rightarrow f^{\alpha}\rightarrow \alpha_{k,\overline{m},\ell}\rightarrow f^s\rightarrow s_{k,\overline{m},\ell}$ with (\ref{equ:MSG_PASS_2})-(\ref{equ:MSG_PASS_3}).
		\STATE Updating the individual MRF according to schemes in Appendix \ref{sec:Appen MRF} with $I_{\rm mrf}$ iterations. Output message $s_{k,\overline{m},\ell}\rightarrow f^s$  with \eqref{equ:MSG_PASS_50}.
		\STATE Message passing over $f^s\rightarrow s_{{\rm c},m,\ell}$ with (\ref{equ:MSG_PASS_5}). Updating the common MRF according to schemes in Appendix \ref{sec:Appen MRF} with $I_{\rm mrf}$ iterations. Output message $s_{{\rm c},\overline{m},\ell}\rightarrow f^s$ with \eqref{equ:MSG_PASS_50_c}.
		\STATE Message passing over $f^s\rightarrow \alpha_{k,\overline{m},\ell}\rightarrow f^\alpha\rightarrow x^{\rm (B)}_{k,m,\ell}$ with \eqref{equ:MSG_PASS_8}-\eqref{equ:MSG_PASS_9}.
		\STATE Update the posterior mean $\hat{{\bf x}}^{\rm (B),pos}_{k,m}$  and posterior variance $\hat{C}_{k,m,\ell}^{\rm (B),pos}$ in Module B with \eqref{equ:module_B_post_mean}- \eqref{equ:module_B_post_var}. \\
		\STATE Update the extrinsic mean $\boldsymbol{\mu}_{k,m}^{\rm (A),pri}$ and variance $\gamma_{k,m}^{\rm (A),pri}$ according to (\ref{equ:ext_B_mean}) and (\ref{equ:ext_B_var}).
		\IF{$iter < I_{\text{max}}$}
		\STATE $iter = iter + 1$.
           \STATE Update parameters $\eta_k$ using \eqref{equ:update_eta_EM} and $\sigma^2_{k,\overline{m},\ell}$ using \eqref{equ:update_sigma_EM}.
           \STATE Update parameters $\omega_{\mathcal{T},1}$ and $\omega_{\mathcal{T},2}$ if $iter\mod ~3==0$  {(for lower complexity)}.
		\ENDIF
		\ENDFOR
		\RETURN The $\hat{{\bf h}}_{k,m} = {\bf F}_{{\rm D},k}\hat{{\bf x}}_{k,m}^{\rm (B)}$ for $\forall k,m$, and $\hat{{\bf H}}_k = [\hat{{\bf h}}_{k,1},\dots,\hat{{\bf h}}_{k,M}].$
	\end{algorithmic}
\end{algorithm}
\section{Frequency Division Pilot Pattern Optimization}\label{sec:pilot_opt}
This section introduces a pilot pattern optimization scheme, to minimize the effect of delay leakage caused by frequency selection in the proposed N-FD-CDM scheme. We begin by re-expressing the sensing matrix for group $g$ -- denoted by $\widetilde{\mathbf{A}}_g,~\forall g$ -- as a function of the pilot selection pattern $\overline{\mathbf{b}}_g$ as follows (referring to \eqref{equ:NFD_CDM},\eqref{equ:H_gkm_to_X}, and \eqref{equ:rx_sig_model_6})
\begin{align}
    \label{eq_sensingA_g}\widetilde{\mathbf{A}}_g(\overline{\mathbf{b}}_g)=&\sqrt{\frac{P^{\rm Tr}}{\mathbf{1}^{\mathrm{T}}\overline{\mathbf{b}}_g}}\mathrm{diag}(\overline{\mathbf{b}}_g)\widetilde{\mathbf{F}}_{{\rm P},g},\end{align}
with 
\begin{align}\notag
\widetilde{\mathbf{F}}_{{\rm P},g}&=\Big[\mathrm{diag}(\widetilde{\mathbf{b}}_{g,1})\mathbf{F}_{{\rm D},k_1},~\cdots,~\mathrm{diag}(\widetilde{\mathbf{b}}_{g,|\mathcal{K}_g|})\mathbf{F}_{{\rm D},k_{|\mathcal{K}_g|}}\Big]\\\notag
&\in\mathbb{C}^{N\times L|\mathcal{K}_g|},
\end{align}
where $k_{i}$ for $i=1,~\cdots,~|\mathcal{K}_g|$ refers to the $i$-th user index in the user set $\mathcal{K}_g$. The term $\sqrt{\frac{P^{\rm Tr}}{\mathbf{1}^{\mathrm{T}}\overline{\mathbf{b}}_g}}$ guarantees that each user transmits the same pilot power of  $P^{\rm Tr}$, regardless of the varying effective pilot-subcarrier lengths among different groups.

Additionally, the MSE of the LMMSE estimator in \eqref{equ:module_A_post_var} only accounts for interference from intra-group users, based on the assumption of orthogonal pilot patterns between groups. However, this assumption cannot be taken for granted when formulating the pilot pattern optimization, where we must involve inter-group interference in the objective MSE formula to promote orthogonal pilots.  Consequently,  the discrete optimization problem w.r.t $\overline{\mathbf{B}}=\big[\overline{\mathbf{b}}_1^{\mathrm{T}},~\cdots,~\overline{\mathbf{b}}_G^{\mathrm{T}}\big]^{\mathrm{T}}\in\{0,1\}^{N\times G}$, to minimize the MSE of the LMMSE estimator in \eqref{equ:module_A_post_var} is formulated as
\begin{maxi!}
        {\overline{\mathbf{B}}}{\sum_{g} {\rm Tr}\Big[\Big( {\lambda}_0^{-1}\mathbf{I}_{L|\mathcal{K}_g|}+\widetilde{\mathbf{A}}_g(\overline{\mathbf{b}}_g)^H\boldsymbol{\Sigma}_g^{-1}\widetilde{\mathbf{A}}_g(\overline{\mathbf{b}}_g)\Big)^{-1}\Big],}{\label{eq_optimization_P1}}{\label{eq_optimization_P1_1}}
        \addConstraint{\overline{B}_{n,g}=\{0,~1\},~\forall n,~g}\label{eq_optimization_P1_2}
        \addConstraint{\mathbf{\overline{B}}\mathbf{1}_G\leq \mathbf{1}_N},\label{eq_optimization_P1_3}
      \end{maxi!}
where $\lambda_0$ denotes the initialized prior variance for the LMMSE estimator in \eqref{equ:module_A_post_var}, specifically given by $\lambda_0=\frac{N}{L}$.  The term $\boldsymbol{\Sigma}_g=\sigma_z^{2}\mathbf{I}_{L|\mathcal{K}_g|}+\lambda_0\sum_{g'}^{g^-}\widetilde{\mathbf{A}}_{g'}(\overline{\mathbf{b}}_{g'})\widetilde{\mathbf{A}}_{g'}^H(\overline{\mathbf{b}}_{g'})$ represents the  noise-plus-interference matrix. \eqref{eq_optimization_P1_1} is a straightforward extension from \eqref{equ:module_A_post_var} using the  {Woodbury Identity \cite{higham2002accuracy}}, taking inter-group interference into account. \eqref{eq_optimization_P1_2} yields the discrete pilot pattern optimization, while \eqref{eq_optimization_P1_3} ensures orthogonality of the pilots between groups.

To relax the discrete constraint of \eqref{eq_optimization_P1_2}, we substitute $\overline{B}_{n,g}$ by a continuous variable $\overline{C}_{n,g}\leq 1$ with the following relationship  $\overline{C}_{n,g}=\sqrt{\overline{B}_{n,g}}$. When substituting the optimization variable in \eqref{eq_optimization_P1} by $\overline{C}_{n,g}$, the resultant $\overline{B}_{n,g}$, after sqrt function, naturally approaches $0$ or $1$. Towards this end, the reformulated optimization becomes
\begin{maxi!}
        {\overline{\mathbf{C}}}{\sum_{g} {\rm Tr}\Big[\Big( {\lambda}_0^{-1}\mathbf{I}_{L|\mathcal{K}_g|}+\widetilde{\mathbf{A}}_g(\overline{\mathbf{c}}^2_g)^H\boldsymbol{\Sigma}_g^{-1}\widetilde{\mathbf{A}}_g(\overline{\mathbf{c}}^2_g)\Big)^{-1}\Big],}{\label{eq_optimization_P2}}{\label{eq_optimization_P2_1}}
        \addConstraint{\overline{C}_{n,g}\leq 1,~\forall n,~g}\label{eq_optimization_P2_2}
        \addConstraint{\mathbf{\overline{C}}\mathbf{1}_G\leq \mathbf{1}_N},\label{eq_optimization_P2_3}
      \end{maxi!}
which can be directly solved by using Adam optimizer \cite{stoica2005spectral} in PyTorch with learning rate $0.1$.  {In \eqref{eq_optimization_P2}, $\overline{\mathbf{c}}^{2}_g$ refers to the vector whose each entry is the square of $\overline{\mathbf{c}}_g$.}

\section{Simulation Results}
In this section, we evaluate the performance of the proposed N-FD-CDM pilot scheme and the proposed Turbo-MRF algorithm, under COST 2100 channel model. In the simulations, we consider $M=1024$ antennas, which are divided into $\overline{M}=128$ sub-arrays with sub-array spacing being $0.21$ m ($5$ times of the half-wavelength) along the horizontal direction. Each sub-array contains $\widetilde{M}=8$ antennas along the vertical direction with half-space wavelength. We consider an OFDM scheme with $1024$ sub-carriers and $72$ cyclic prefix length.  {The sub-carriers allocated to each frequency group are determined by the pilot pattern optimization in Section V.} The central frequency is at $3.5$ GHz, with the frequency spacing being $15$ kHz. 

As for the users, we randomly generate their locations uniformly within two circles centering at $[20,10,1.5]$ m and $[-20,-18,1.5]$ m (the round-trip delay is guaranteed to be smaller than OFDM's guard interval) with the diameter of $10$ m.  {To ensure low interference between users, we group the users according to a rule that maximizes the sum of intra-group user distance.}   

Throughout simulations, the average channel power of each user is normalized to $1$, and the power of each user is defined as $P^{\rm Tr}_k=\|\mathbf{u}_k\|^2,~\forall ~k$. Then we define the SNR per user as
\begin{align}
    \label{equ:SNR}
    {\rm SNR}_k=\frac{P^{\rm Tr}_k}{N \sigma_z^2},
\end{align}
which is set the same across users.

 We evaluate the normalize MSE (NMSE) of the channel estimation performance, which is given by
\begin{align}
    \label{equ:NMSE}
    {\rm NMSE}=\frac{\sum_k\|{\mathbf{H}}_k-\hat{\mathbf{H}}_k\|_{\rm F}^2}{\sum_k\|{\mathbf{H}}_k\|_{\rm F}^2},
\end{align}
where $\hat{\mathbf{H}}_k$ is the estimated channel frequency response.

 Other algorithm and pilot scheme baselines are considered for performance evaluation as follows.
\begin{itemize}
    \item \textbf{LMMSE}: A fundamental baseline of channel estimation method without exploiting the sparsity prior, which adopts \eqref{equ:module_A_post_mean} by setting $\boldsymbol{\mu}^{\rm (A),pri}_{k,m}=\mathbf{0}$ and ${\gamma}^{\rm (A),pri}_{k,m}=N/L$. 
    \item \textbf{LMMSE-genie}: The best-performance (but almost unachievable) baseline with prior knowledge of each user's exact channel power at each delay tap of each antenna. Specifically, it adopts  {\eqref{equ:module_A_post_mean} } by  {setting $\boldsymbol{\mu}^{\rm (A),pri}_{k,m}=\mathbf{0}$ and ${\gamma}^{\rm (A),pri}_{k,m,\ell}=|X_{k,m,\ell}|^2$ (In this case,  ${\gamma}^{\rm (A),pri}_{k,m,\ell}$ has an additional sub-script $\ell$ as we do not assume uniform variance across the delay taps for the best NMSE lower bound.)}.
    \item \textbf{VAMP-BG}: An algorithm baseline that employs our proposed N-FD-CDM pilot scheme but does not incorporate the MRF prior, as utilized in our proposed Turbo-MRF method. The VAMP-BG assumes i.i.d Bernoulli Gaussian prior \cite{VAMP}.
    \item  {\textbf{OMP}: An algorithm baseline that  uses orthogonal matching pursuit (OMP)-based compressed channel estimation for channel estimation \cite{8316247}.}
    \item \textbf{CDM}:  {A pilot scheme baseline using CDM\footnote{For complexity fairness and tractability, the channel estimation algorithm applies individual LMMSE updated with message passing.} with random-phase pilot \cite{1542412}.} 
    
     \item \textbf{SR-FDM}: A pilot scheme baseline using FDM \cite{bossert2002cyclic}, which uses $1$ OFDM symbols for channel estimation for all users. In this scheme, the pilot patterns among different frequency groups are periodically intersected.  The channel estimation applies VAMP-BG.
     
     \item  {\textbf{NR Orthogonal}: Another  pilot scheme baseline using FDM, which serves at most $8$  users per OFDM symbol according to the standard NR practices \cite{3gpp_38_211}, ensuring orthogonality for the channel estimation algorithm. This pilot scheme serves as a fundamental baseline to demonstrate the pilot reduction ratio of our proposed pilot and channel estimation scheme.}

\end{itemize}

\begin{figure}[t]
    \centering
    \includegraphics[width=0.95\linewidth]{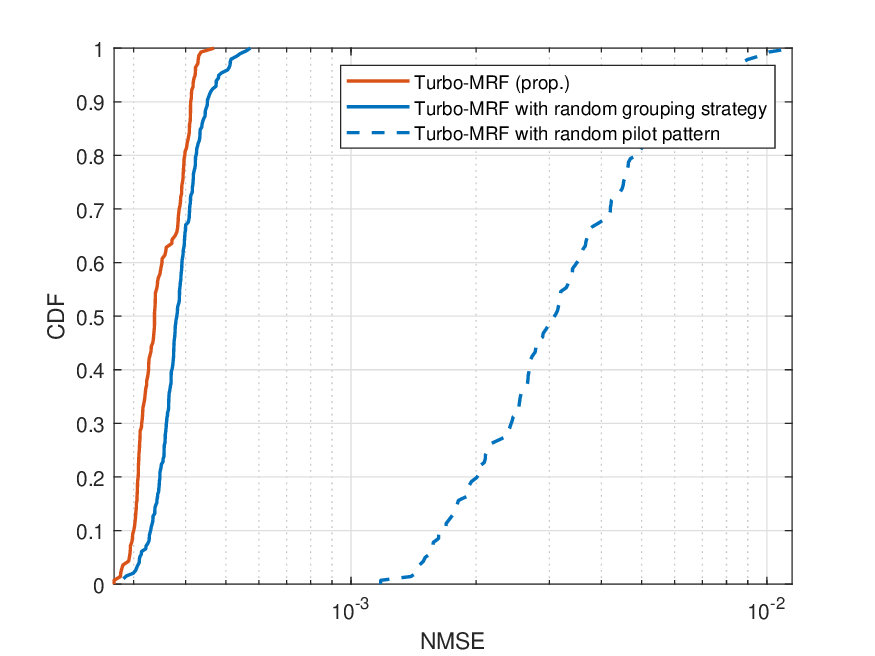}
    \caption{ {The CDF plot for $24$ users under $20$ dB SNR, comparing between the proposed Turbo-MRF  scheme (solid-red, with the proposed grouping strategy and the optimized pilot pattern), the Turbo-MRF with random grouping strategy (solid-blue, with optimized pilot patterns), and the Turbo-MRF with random pilot patterns (dashed-blue, with proposed grouping strategy).}}
    \label{fig:CDF}
\end{figure}

 {We first verify the effectiveness of our optimized pilot pattern in Section \ref{sec:pilot_opt} and grouping strategy in Remark 1 through Fig. \ref{fig:CDF}. Specifically, Fig. \ref{fig:CDF} compares the cumulative distribution function (CDF) of channel estimation performance between our proposed Turbo-MRF (N-FD-CDM) scheme (solid red, incorporating both the proposed grouping strategy and optimized pilot pattern) and two benchmark scenarios: Turbo-MRF with a random grouping strategy (solid blue) and Turbo-MRF with random pilot patterns (dashed blue). Firstly, the results in Fig. \ref{fig:CDF} demonstrate that the proposed grouping strategy efficiently enhances the channel estimation performance, thereby validating the advantages of leveraging the spatial CIR support prior through the MRF approach.} Moreover, Fig. \ref{fig:CDF} highlights that, without the optimized pilot pattern proposed in Section \ref{sec:pilot_opt}, the robustness of channel estimation performance  in multi-user XL-MIMO systems significantly degrades. This variability is expected, as given a large number of user groups, random pilot selection can result in poor pilot patterns for at least one frequency group with high probability. Therefore, in the following simulations, we focus on channel estimation performance with optimized pilot patterns and proposed grouping strategy.

\begin{figure}[!t]
\centering
\begin{subfigure}[t]{0.5\textwidth}{
    \begin{minipage}{0.95\textwidth}
	\includegraphics[width=\textwidth]{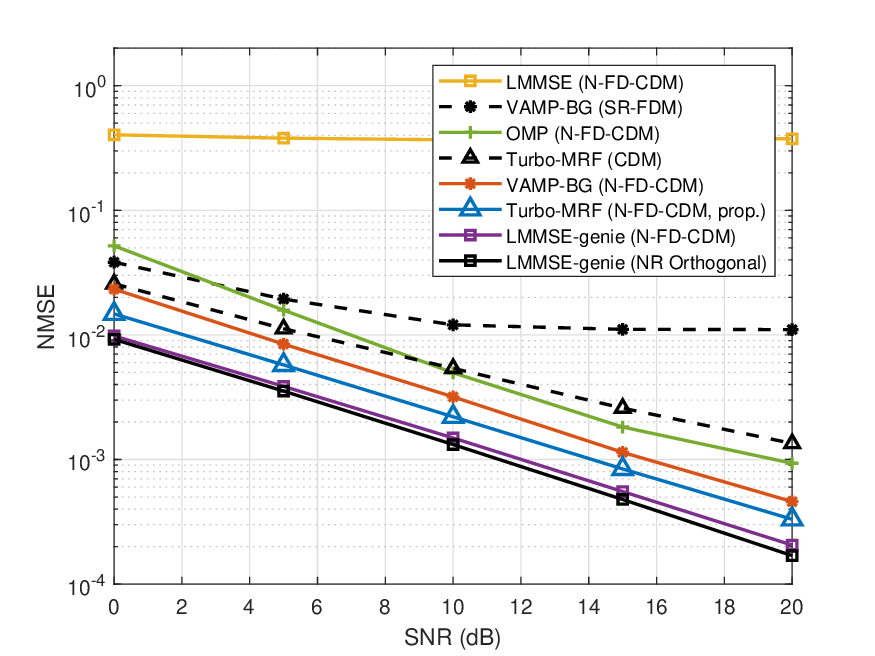}
	\caption{$K=24$}\label{fig:SNR_K24}
    \end{minipage}
	}
\end{subfigure}
\begin{subfigure}[t]{0.5\textwidth}{
    \begin{minipage}{0.95\textwidth}
	\includegraphics[width=\textwidth]{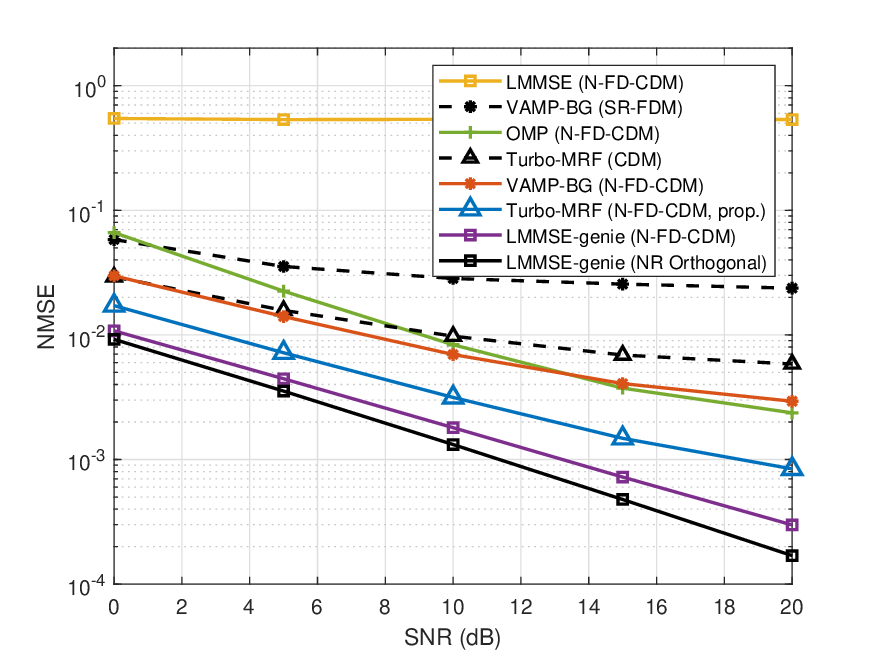}
	\caption{$K=32$}\label{fig:SNR_K32}
    \end{minipage}
	}
\end{subfigure}
\begin{subfigure}[t]{0.5\textwidth}{
    \begin{minipage}{0.95\textwidth}
	\includegraphics[width=\textwidth]{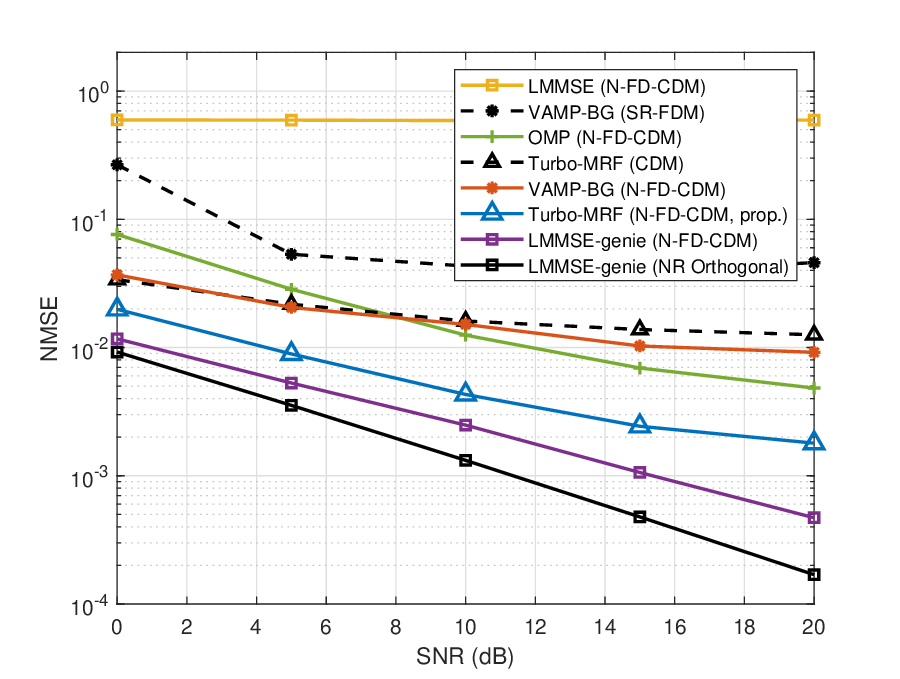}
	\caption{$K=36$}\label{fig:SNR_K36}
    \end{minipage}
	}
\end{subfigure}
	\caption{ {NMSE performance of the multi-user XL-MIMO uplink channel estimation as a function of SNR, comparing different numbers of users.}}
	\label{fig:SNR}
\end{figure}

Then, we present the NMSE as a function of SNR for varying numbers of users, as illustrated in Fig. \ref{fig:SNR}. In all cases, the proposed Turbo-MRF (N-FD-CDM)  demonstrates the best performance, except when compared to the LMMSE-genie method, which assumes known ground-truth power of the CIRs. Notably, the LMMSE channel estimation method performs the worst even with the novel pilot scheme, as it does not exploit any sparsity properties.

More importantly,  {Fig. \ref{fig:SNR} indicates that the proposed Turbo-MRF (N-FD-CDM) scheme supports the maximum number of users, by achieving the lowest NMSE given an arbitrary number of users (except for the pilots schemes using LMMSE-genie).} In addition,  {Fig. \ref{fig:SNR}(\subref{fig:SNR_K24}) shows that, adopting the LMMSE-genie algorithm, our proposed N-FD-CDM pilot scheme achieves performance nearly equivalent to that of the NR orthogonal pilot scheme with near-field ELAA, despite using only $1$ OFDM symbol compared to the $3$ OFDM symbols required by the NR orthogonal pilot scheme. Therefore, we can conclude that our proposed N-FD-CDM pilot scheme effectively saves $2/3$ of the overhead with near-field ELAA.}

 {In addition, Fig. \ref{fig:SNR}(\subref{fig:SNR_K24}) indicates that the traditional VAMP-BG (SR-FDM), despite utilizing a sparsity recovery algorithm, reach an error floor of around $10^{-2}$ for $24$ users, regardless of increasing SNR.} Similarly, although applied with our proposed Turbo-MRF algorithm, the CDM pilot scheme tends to saturate around $6 * 10^{-3}$ for $32$ users,  {while our proposed N-FD-CDM pilot scheme, incorporating with compressed sensing algorithms (VAMP-BG and OMP), saturates around $3*10^{-3}$.} 
Notably, our proposed Turbo-MRF (N-FD-CDM) scheme overcomes this error floor, whose  NMSE performance continues to improve with increasing SNR, exceeding beyond $0.1 \times 10^{-3}$ at $20 $ dB SNR. At $K=36$ in Fig. \ref{fig:SNR}(\subref{fig:SNR_K36}), all algorithms exhibit an error floor; however, the proposed Turbo-MRF (N-FD-CDM) shows superior precision, achieving below $2*10^{-3}$ at $20$ dB SNR compared to over $5*10^{-3}$ for  {Turbo-MRF (CDM), VAMP-BG (N-FD-CDM) and OMP (N-FD-CDM)}.

\begin{figure}[t]
    \centering
    \includegraphics[width=1\linewidth]{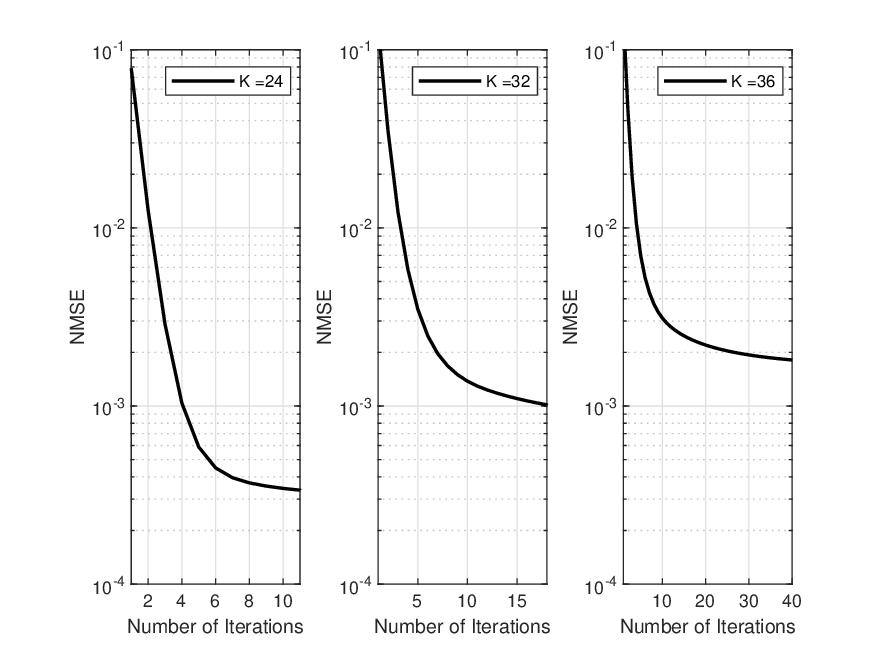}
    \caption{ {The convergence plot of the proposed Turbo-MRF scheme for Figs. \ref{fig:SNR}.}}
    \label{fig:convergence}
\end{figure}

 {The proposed Turbo-MRF algorithm is a heuristic algorithm whose convergence is hard to be theoretically validated. However, this could be illustrated in the convergence plot as shown in Fig. \ref{fig:convergence}. Generally, for different user scenarios, the proposed Turbo-MRF algorithm converges within 20 iterations.}

\begin{figure}[t]
    \centering
    \includegraphics[width=1\linewidth]{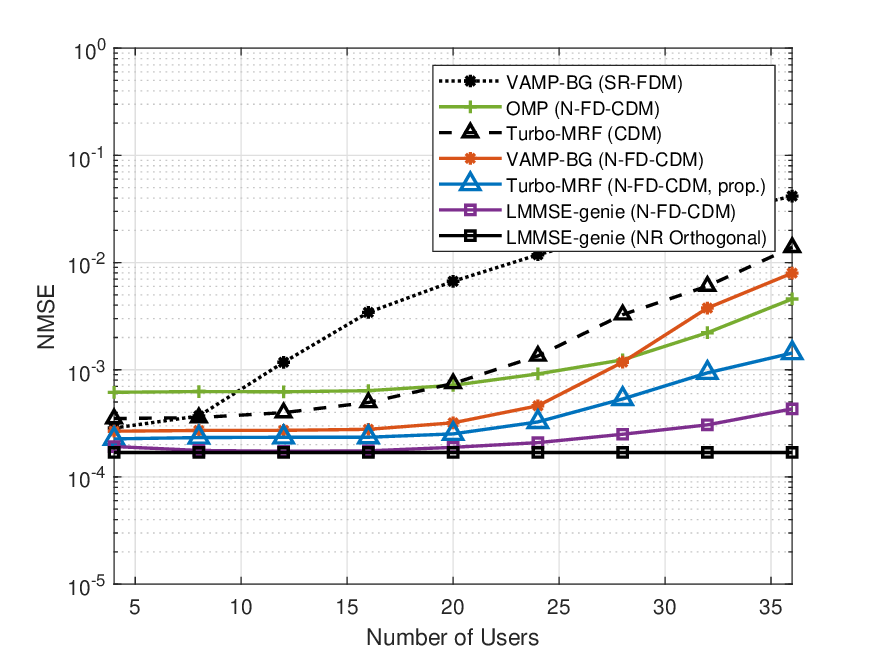}
    \caption{NMSE performance as a function of the number of users under $20$ dB SNR.}
    \label{fig:diff_K}
\end{figure}

To further assess the performance of the N-FD-CDM pilot scheme, we plot the NMSE as a function of $K$ in Fig. \ref{fig:diff_K}.  {The results indicate that our proposed Turbo-MRF (N-FD-CDM) scheme maintains a consistent  NMSE performance that approaches the NMSE performance of NR orthogonal pilot for up to $24$ users.} Beyond this point, as the allocated sub-carriers per group drop below $72$, significant performance degradation occurs.  {Nevertheless, for $K \geq 24$, the performance advantage of Turbo-MRF (N-FD-CDM) over VAMP-BG (N-FD-CDM, Turbo-MRF (CDM) and OMP (N-FD-CDM) is still significant, reaching over $3$ dB for $K\geq 28$.} From Fig. \ref{fig:diff_K}, it is evident that to ensure an NMSE of $10^{-3} $, the proposed Turbo-MRF (N-FD-CDM) scheme can serve at least $32$ users, while the VAMP-BG (N-FD-CDM)/OMP (N-FD-CDM) methods can only accommodate around $28/20$ users, all exceeding the traditional VAMP-BG (SR-FDM) scheme's capacity of approximately $12$ users and the Turbo-MRF (CDM) method's capacity of $20$ users. Thus, the proposed N-FD-CDM enables the XL-MIMO system to support more users. 

\begin{figure}[t]
    \centering
    \includegraphics[width=0.95\linewidth]{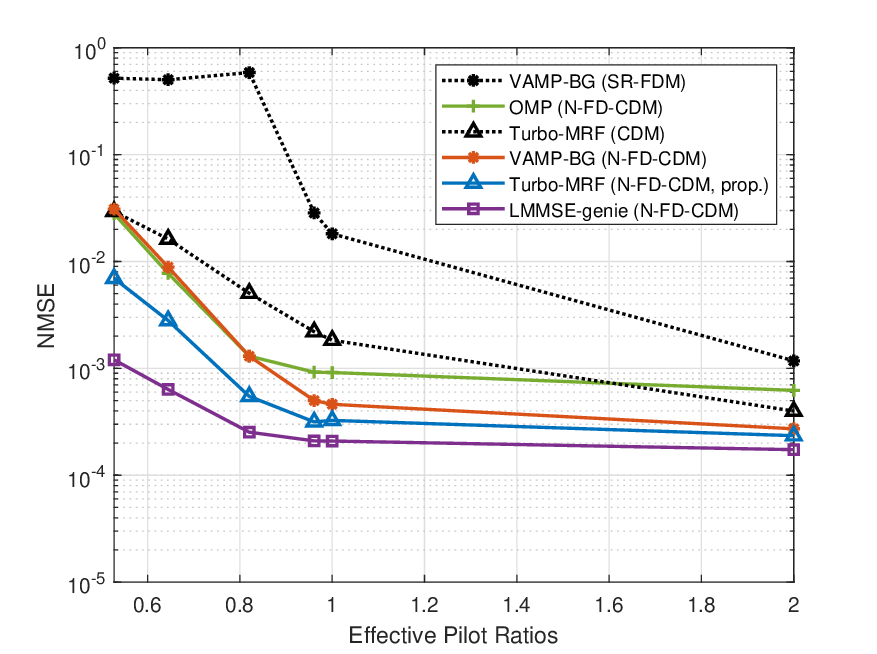}
    \caption{ {NMSE performance as a function of the effective pilot ratio for $24$ users under $20$ dB SNR.}}
    \label{fig:diff_pilot_ratio}
\end{figure}

 {To further demonstrate the capability of reducing pilot overhead in our proposed Turbo-MRF (N-FD-CDM) scheme, we plot the NMSE as a function of the effective pilot ratio (the number of effective pilots over the total number of pilots) in Fig. \ref{fig:diff_pilot_ratio}. The results show that to guarantee an NMSE of  $10^{-3} $, the proposed Turbo-MRF  (N-FD-CDM) scheme requires at most $85\%$ pilot overhead, while the VAMP-BG (N-FD-CDM) and OMP (N-FD-CDM) methods require at least $95\%$, which is unachievable for the Turbo-MRF  (CDM) and VAMP-BG (SR-FDM) methods given only $1$ OFDM symbol. This further illustrates the ability of our proposed Turbo-MRF scheme to support more users under the same hardware conditions in XL-MIMO.}

\section{Conclusions}
This paper introduced a novel pilot scheme, termed N-FD-CDM,  to enhance uplink channel sounding performance for multi-user OFDM XL-MIMO systems, with a sub-array structured ELAA. The N-FD-CDM pilot scheme fully exploits the clustered sparsity of users' CIR to reduce pilot overheads. Additionally, we propose a channel estimation algorithm called Turbo-MRF, which is based on a Bayesian inference framework with a 2-D MRF prior that effectively captures the clustered sparsity of the CIR. Simulations demonstrate that our proposed N-FD-CDM pilot scheme, in conjunction with the Turbo-MRF algorithm, enables channel estimation for a greater number of users in XL-MIMO systems.

\section{Acknowledgment}
The authors would like to thank Nilesh Kumar Jha for his baseline codes on orthogonal pilots.
\appendix

\subsection{Message passing over MRF}\label{sec:Appen MRF}
When feeding $\pi_{k,\overline{m},\ell}^{\text{in},s}$ (or $\pi_{{\rm c},\overline{m},\ell}^{\text{in},s}$) into each individual MRF sub-graph (or common MRF sub-graph), the message passing can be performed over four directions, i.e., from left to right, from right to left, from bottom to top and from top to bottom, whose passing messages are denoted as  $\lambda_{k,\overline{m},\ell}^\mathcal{D}$ with $\mathcal{D}=\{\rm L,R,T,B\}$ respectively. For convergence, we update $\lambda_{k,\overline{m},\ell}^L$, $\lambda_{k,\overline{m},\ell}^R$, $\lambda_{k,\overline{m},\ell}^{\mathrm{T}}$ and $\lambda_{k,\overline{m},\ell}^B$ sequentially over $I_{\rm mrf}$ loops \cite{Yuan_MRF}. Consequently, the MRF updating scheme can be summarized as follows
    \begin{itemize}
        \item   {Initialization} $\lambda_{k,\overline{m},\ell}^{\mathcal{D}}=0.5, ~\forall ~\overline{m},~\ell $; $I_{\rm mrf}$
        \item  {for $i\leq I_{\rm mrf}$:}
        \begin{itemize}
            \item update $\lambda_{k,\overline{m},\ell}^{\rm {L}}, \forall \overline{m},\ell$  from \eqref{equ:MRF_msg_neighbor} (We take $\mathcal{D}=\rm L$ as an example here and $\eta_{\lambda}\in\{\rm L,T,B\}$.);
             \item update $\lambda_{k,\overline{m},\ell}^{\rm {R}}, \forall \overline{m},\ell$, $\lambda_{k,\overline{m},\ell}^{\rm {T}}, \forall \overline{m},\ell$ and  $\lambda_{k,\overline{m},\ell}^{\rm {B}}, \forall \overline{m},\ell$ sequentially similarly to \eqref{equ:MRF_msg_neighbor};
        \end{itemize}
        \item  Update $\pi_{k,\overline{m},\ell}^{\text{out},s}$ with (\ref{equ:MRF_OutPut_Individual}) for individual MRF output and $\pi_{k,\overline{m},\ell}^{\text{out},s_c}$ in \eqref{equ:MRF_OutPut_common} for common MRF output.
    \end{itemize}

\begin{figure*}[!tbh]
\begin{align}
	\label{equ:MRF_msg_neighbor}
	\lambda_{k,\overline{m},\ell}^{L}= &\frac{\pi_{\mathcal{T},\overline{m},\ell-1}^{\text{in},s}\prod_\eta \lambda_{k,\overline{m},\ell-1}^{\eta}e^{w_1-w_2}+\left(1-\pi_{\mathcal{T},\overline{m},\ell-1}^{\text{in},s}\right)\prod_\eta \left(1-\lambda_{k,\overline{m},\ell-1}^{\eta}\right)e^{w_2-w_1}}{\left(e^{w_1}+e^{-w_1}\right)\left(\pi_{\mathcal{T},\overline{m},\ell-1}^{\text{in},s}\prod_\eta \lambda_{k,\overline{m},\ell-1}^{\eta}e^{-w_2}+\left(1-\pi_{\mathcal{T},\overline{m},\ell-1}^{\text{in},s}\right)\prod_\eta \left(1-\lambda_{k,\overline{m},\ell-1}^{\eta}\right)e^{w_2}\right)},\\
    \label{equ:MRF_OutPut_Individual}
    \pi_{k,\overline{m},\ell}^{\text{out},s}=& \frac{e^{-w_2/2}\prod_{\eta}\lambda_{k,\overline{m},\ell}^{\eta}}{e^{-w_2/2}\prod_{\eta}\lambda_{k,\overline{m},\ell}^{\eta} + e^{w_2/2}\prod_{\eta}\left(1-\lambda_{k,\overline{m},\ell}^{\eta}\right)},~\eta\in\{\rm L,R,T,B\},\\
     \label{equ:MRF_OutPut_common}
    \pi_{k,\overline{m},\ell}^{\text{out},s_{\rm c}}=& \left(1+\frac{ e^{w_2/2}\left[\prod_{\eta}\left(1-\lambda_{k,\overline{m},\ell}^{\eta}\right)\right]\left(1-\pi^{{\rm in},s}_{c,\overline{m},\ell}\right)/\left(1-\pi^{{\rm in},s_{\rm c}}_{k,\overline{m},\ell}\right)}{e^{-w_2/2}(\prod_{\eta}\lambda_{k,\overline{m},\ell}^{\eta})\pi^{{\rm in},s}_{c,\overline{m},\ell}/\pi^{{\rm in},s_{\rm c}}_{k,\overline{m},\ell}}\right)^{-1},~\eta\in\{\rm L,R,T,B\}
\end{align}
	\hrulefill
\end{figure*}

\subsection{EM updates}\label{sec:Appen EM}
 {According to \cite{vila2011expectation}, the EM algorithm updates each parameter iteratively by maximizing the posterior expectation of the log-likelihood function of the prior distribution in \eqref{equ:overall_model_pri}.    Specifically, we select $\big\{\{{\bf X}_k\},\{{\boldsymbol{\alpha}}_{k,\overline{m}}\},\{{\bf S}_k\},{\bf S}_{\rm c}\big\}$ to be the hidden variables, which yields the following EM update rule}
\begin{align}  
    \label{equ:EM update}
    \notag\mathbf{q}^{t+1}=&  \arg\max_{\bf q} ~\mathbb{E}\Big\{ \sum_{k,\overline{m}}\log~ p\left(\overline{{\bf X}}_{k,\overline{m}}|\boldsymbol{\alpha}_{k,\overline{m}}\right)+\\\notag&  \sum_{k,\overline{m}} \log~ p\left(\boldsymbol{\alpha}_{k,\overline{m}}|{\bf s}_{k,\overline{m}},{\bf s}_{{\rm c},\overline{m}}\right)+\sum_k \log~ p({\bf S}_k;\mathcal{W}_k)+\\&   \log~ p({\bf S}_{\rm c};\mathcal{W}_{\rm c})\Big|\mathbf{q}^{t} \Big\}.
\end{align}
 {For low complexity, we solve \eqref{equ:EM update} by updating each individual parameter while fixing the others, iteratively.}

\begin{lem}\label{lem:EM}
      {Following EM, at the $(t+1)$-th iteration, the parameters are updated by}
    \begin{align}  
        \label{equ:update_eta_EM_app}
        \eta^{t+1}_k=&    \arg \max_{0\leq\eta_k\leq 1} ~\mathbb{E}\Big\{\sum_{k,\overline{m}}\log ~p\left(\boldsymbol{\alpha}_{k,\overline{m}}|{\bf s}_{k,\overline{m}},{\bf s}_{{\rm c},\overline{m}}\right) \Big|\mathbf{q}^{t}\Big\},\\
        \label{equ:update_sigma_EM_app}  \sigma^{2,t+1}_{k,\overline{m},\ell}=&  \arg \max_{\sigma_{k,\overline{m},\ell}>0}  \mathbb{E}\big\{\sum_{k,\overline{m}}\log~ p\left({\bf x}_{k,\overline{m}}|{\boldsymbol{\alpha}}_{k,\overline{m}}\right) \Big|\mathbf{q}^{t}\big\},\\
         \label{equ:update_w1}
       \omega_{\mathcal{T},1}^{t+1}+=&  \Delta \sum_{\overline{m},\ell}m_{\mathcal{G},\overline{m},\ell}\left[p^{{\rm pos}}_{\mathcal{T},\overline{m},\ell}-\tanh(\omega_{\mathcal{T},1}m_{\mathcal{T},\overline{m},\ell}+\omega_{\mathcal{T},2})\right],\\
      \label{equ:update_w2}  
     \omega_{\mathcal{T},2}^{t+1}+=  &  \Delta \sum_{\overline{m},\ell}\left[p^{{\rm pos}}_{\mathcal{T},\overline{m},\ell}-\tanh(\omega_{\mathcal{T},1}m_{\mathcal{T},\overline{m},\ell}+\omega_{\mathcal{T},2})\right],
    \end{align}
    with
    \begin{align}
          \notag p^{{\rm pos}}_{\mathcal{T},\overline{m},\ell}\propto &  \nu_{s_{\mathcal{T},\overline{m},\ell}\rightarrow f^s}(s_{\mathcal{T},\overline{m},\ell})\nu_{ f^{s}\rightarrow s_{\mathcal{T},\overline{m},\ell}}(s_{\mathcal{T},\overline{m},\ell}),~\forall k,\\
        \notag    m_{\mathcal{T},\overline{m},\ell}=&  \sum_{\overline{m}',\ell'\in\mathcal{I}_{\overline{m},\ell}}p^{\rm pos}_{\mathcal{T},\overline{m},\ell}p^{\rm pos}_{\mathcal{T},\overline{m}',\ell'},
    \end{align}
     {where \eqref{equ:update_w1} and \eqref{equ:update_w2} in Lemma \ref{lem:EM} refers to a gradient-descent update with step-size $\Delta$ and $^{\rm pos}_{\mathcal{T},\overline{m},\ell}$ is short for $p(s_{\mathcal{T},\overline{m},\ell}=1\big|\mathbf{q}^{t})$.  $a+=b$ represents $a\leftarrow a+b$.}
\end{lem}
\begin{IEEEproof}
 {\eqref{equ:update_eta_EM_app} and \eqref{equ:update_sigma_EM_app} in Lemma \ref{lem:EM} result in \eqref{equ:update_eta_EM} and \eqref{equ:update_sigma_EM} respectively, and are straightforward by selecting the involved variables setting the first-order derivative equal to $0$. \eqref{equ:update_w1} and \eqref{equ:update_w2} in Lemma \ref{lem:EM} refers to a gradient-descent method of solving the pseudo-likelihood expression of  \eqref{equ:EM update} referring to  \cite[Result 1.1]{ghosal2020joint}, as its original form is computationally intractable. }
\end{IEEEproof}

\ifCLASSOPTIONcaptionsoff
  \newpage
\fi
\bibliographystyle{IEEEtran}
\bibliography{Main.bib}

\end{document}

%% file: notation_new.tex
\def\nb0{{\mathbf{0}}}
\def\nb1{{\mathbf{1}}}

\newtheorem{lem}{Lemma}

\newtheorem{eg}{Example}

